\documentclass[11pt,
a4paper,
oneside,
notitlepage 
]{article}

\usepackage[T1]{fontenc}
\usepackage{jheppub}
\usepackage{amsmath}
\usepackage{amsfonts}
\usepackage{amssymb}
\usepackage{graphicx}
\usepackage{amsthm}
\usepackage{physics}
\usepackage{comment}
\usepackage{bbold}
\usepackage{slashed}
\usepackage{caption}
\usepackage{subcaption}
\usepackage{cases}
\usepackage[toc,page]{appendix}
\usepackage{titlesec}
\usepackage{thmtools}
\usepackage{float}
\usepackage{empheq}
\usepackage[dvipsnames]{xcolor}
\usepackage[many]{tcolorbox}
\usepackage{microtype}
\usepackage{hyperref}
\usepackage{cleveref}

\graphicspath{{./pics}}

\newcommand{\M}[2]{M\begin{pmatrix}#1\\#2\end{pmatrix}}
\newcommand{\G}[2]{G\begin{pmatrix}#1\\#2\end{pmatrix}}
\newcommand{\ddv}[3]{\frac{\dd^2 #1}{\dd #2 \dd #3}}
\newcommand{\dvv}[2]{\frac{\dd^2 #1}{\dd #2^2}}
\newcommand{\lin}{\lambda_{\text{in}}}
\newcommand{\GeV}{\text{ GeV}}

\declaretheoremstyle[headfont=\bfseries\scshape,postheadspace=\newline]{mystyle}

\declaretheorem[style=mystyle]{convention}
\declaretheorem[style=mystyle]{assumption}
\declaretheorem[style=mystyle]{theorem}
\declaretheorem[style=mystyle]{definition}
\declaretheorem[style=mystyle]{lemma}
\declaretheorem[style=mystyle]{remark}
\declaretheorem[style=mystyle]{corollary}
\declaretheorem[style=mystyle]{property}

\begin{document}

\title{A Scattering Phase Space Primer}
\author[a]{Francesca Acanfora}
\emailAdd{facanfora@umass.edu}
\affiliation[a]{Department of Physics, University of Massachusetts Amherst, MA 01003, USA}
\date{}

\abstract{
The kinematic phase space is fundamental to all particle scattering processes across experiments - from lepton and hadron colliders to dark matter searches. While $2 \to 2$ scatterings are well understood, the simplest non-trivial case of three-body final states presents major analytic challenges that are typically circumvented by numerical methods. This paper provides the first comprehensive analytic treatment of $2 \to 3$ phase space and cross section distributions that uses a non-constant matrix element and keeps all particle masses non-zero. This is fundamental for an accurate description of crucial configurations for Beyond Standard Model searches, e.g. Vector Boson Fusion. Using Gram Determinants as our primary language for Lorentz-invariant kinematics, we derive exact, closed-form expressions for differential cross section distributions and explicitly construct the complete physical phase space boundaries. 
This paper serves as both a pedagogical introduction and a practical reference for the analytic computation of realistic $2 \to 3$ processes, showing how the $m\neq0, |{\cal M}|\neq 1$ problem is computationally tractable through proper kinematic methods. The results apply universally across all $2 \to 3$ scatterings regardless of center-of-mass energy, particle masses, or interaction type.
}

\maketitle

\section{Introduction}
There is something that a 100 TeV future hadron collider \cite{fcchh} and a 10 GeV lepton collider \cite{belleii} have in common. It is the same thing that brings together the Higgs discovery \cite{higgs_discovery} with a nuclear recoil in a Dark Matter search \cite{LZ:2026axp}. Under this respect the collision of muon beams \cite{muc,mutristan} is not that different with respect to that of protons dumped on a target \cite{NA62:2017rwk}. The common denominator is the kinematic phase space of a scattering \cite{PDG:2026}. Having a full understanding of the kinematic phase space is at least as important as controlling the details of the theory we want to study, whatever it is. In some sense, we can think of a collider experiment (or of a big tank of Xenon exposed to the Dark Matter wind, etc.) as a sophisticated device for the creation of kinematic phase spaces. In this epoch of great ferment for the creation of future colliders \cite{hllhc,fccee,fcchh,clic,muon_collider,cepc} it is mandatory that we know every detail of the thousands of phases spaces these machines will produce every second of their activity. Indeed, the BSM theories a collider can probe, the Dark Matter models a detection experiment can exclude or the nuclear interaction hypotheses an experiment can test are almost countless, but the underlying kinematic phase space is one. Improving our knowledge of it benefits all of the above analyses.\\
The knowledge accumulated on $2 \to n$ phase spaces is vast, but is often given for granted, scattered across the literature, hidden in specialized or decades-old references, and increasingly obscured by the increasing ease with which modern computational tools can generate and integrate them \cite{madgraph,whizard,heptapod}. 
However knowing basic properties of phase spaces can guide us to the right answer before many complicated and (computationally and economically) expensive simulations. This is particularly true for Vector Boson Fusion scatterings. Their total cross section is bounded only because the masses in the scattering are non zero. The cross section drastically increases when these become particularly small with respect to the center of mass energy, making Monte Carlo event generators unreliable. Using an Equivalent Vector Approximation is an option, but its applicability needs to be verified case by case and is not guaranteed for any value of the masses in the scattering \cite{eva_fail}. 

\begin{figure}[!h]
\centering
\includegraphics[width=0.55\textwidth]{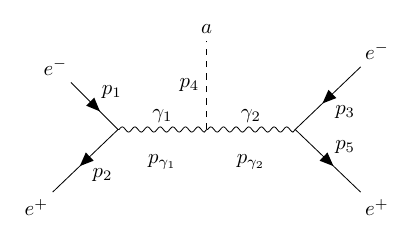}
\includegraphics[width=0.35\textwidth]{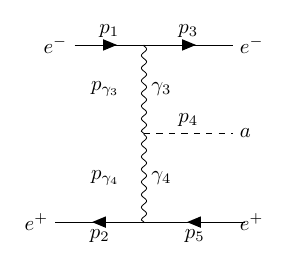}
\caption{$e^+ e^- \to e^+ e^- a$ in an $s-$ and $t-$channel-like topology (left and right panel respectively.}
\label{fig:alp_chans}
\end{figure}

In particular, this work stems from the necessity of analytically computing the cross section of a non trivial, massive $2\to3$ scattering, e.g. $e^+ e^- \to e^+ e^- a$, where $a$ is an Axion-Like Particle (ALP). For this reason throughout this work we will use this scattering as an example. The scattering comes in an ALP-strahlung (\cref{fig:alp_chans}, left panel) and a Vector Boson Fusion topology (\cref{fig:alp_chans}, right panel). The latter would be divergent if the electron mass was identically 0. A phenomenological study of this process at Belle II can be found in \cite{Acanfora:2023gzr} but the results of this work can be applied to any $2\to3$ scattering regardless of the Center of Mass energy, the mass and nature of the particles and their interaction. Indeed a fully scalar theory can be a good approximation to capture the almost divergent nature of VBF. We refer the reader to \cite{gram_epa} for a quantification of the errors we make in different forms of scatterings approximations (neglecting spins, neglecting $s-$channels, neglecting phase space regions where the matrix element is subdominant, a.k.a. Equivalent Vector Approximation).

Resorting to a Lorentz invariant treatment stems from the failure to compute the analytic cross section for $e^+ e^- \to e^+ e^- a$ in polar coordinates:

\begin{equation}
\sigma=\sigma_0 \int \frac{\dd^3 p_3}{(2 \pi)^3 2 E_3}\frac{\dd^3 p_4}{(2 \pi)^3 2 E_4}\frac{\dd^3 p_5}{(2 \pi)^3 2 E_5} (2\pi)^4 \delta^{(4)}(p_1+p_2-p_3-p_4-p_5) |{\cal M}|^2.
\end{equation}
We used the four-momenta naming of \cref{fig:alp_chans} and collected all constants in $\sigma_0$. We can integrate the phase space of $p_4$ against the four-momentum conservation Dirac $\delta$ distribution

\begin{equation}
\sigma=\sigma_0 \int |\vec{p}_3| \dd E_3 \dd c_3 \dd \phi_3 |\vec{p}_5| \dd E_5 \dd c_5 \delta((p_1+p_2-p_3-p_5)^2-m_4^2) |{\cal M}|^2,
\end{equation}
where we explicitly used polar coordinates, we called $\cos \theta_{3,5}=c_{3,5}$, we rotated the azimuthal angle of $p_5$ away and redefined $\sigma_0$. Now

\begin{equation}
\begin{aligned}
(p_1+p_2-p_3-p_5)^2&=s-2\sqrt{s}(E_3+E_5)+m_3^2+m_5^2+2 E_3 E_5+\\
&-2 \sqrt{E_3^2-m_3^2}\sqrt{E_5^2-m_5^2}(s_3 s_5 \cos \phi_3+c_3 c_5)
\end{aligned}
\end{equation}
The easiest way of imposing four-momentum conservation is solving it for $\cos \phi_3$ because the argument of the $\delta$ distribution is linear in it. With the definitions

\begin{align}
&\dd \mu_P:=|\vec{p}_3| \dd E_3 \dd c_3 \dd \phi_3 |\vec{p}_5| \dd E_5 \dd c_5,\label{def:pol_measure}\\
&-b:=s-2\sqrt{s}(E_3+E_5)+m_3^2+m_5^2+2 E_3 E_5-2 \sqrt{E_3^2-m_3^2}\sqrt{E_5^2-m_5^2} c_3 c_5,\\
&a:=-2 \sqrt{E_3^2-m_3^2}\sqrt{E_5^2-m_5^2} s_3 s_5 \cos \phi_3,
\end{align}
the Dirac $\delta$ distribution is solved by

\begin{equation}
\bar{\phi}:=\arccos (b/a).
\end{equation}
The cross section becomes

\begin{equation}
\begin{aligned}
\sigma&=\sigma_0 \int \dd \mu_P \int \frac{\dd \cos \phi_3}{|\sin \phi_3|} \delta(a \cos \phi_3-b) |{\cal M}|^2\\
&=\sigma_0 \int \dd \mu_P |{\cal M}|^2_{\phi_3 \to \bar{\phi}} \frac{\theta\left(\frac{b}{a}+1\right)\theta\left(1-\frac{b}{a}\right)}{|a| \sqrt{1-\left(\dfrac{b}{a}\right)^2}}.\label{eq:dd_jacobian}
\end{aligned}
\end{equation}
The two $\theta$ functions come from imposing $0\leq \phi_3 \leq 2 \pi$. They define this scattering phase space but are highly non trivial to solve. The jacobian adds to the complexity of this cross section.\\ 
Gram Determinants eliminate this complexity. The aim of this work is walking the reader through all the steps to solve \cref{eq:dd_jacobian}. Fundamental to this goal was gathering all the necessary information on the Lorentz invariant phase space scattered across old and cryptic literature. We deliberately organize the presentation around definitions, theorems, corollaries, and proofs so to provide explicit reference points, besides \cref{app:definitions}, to which the reader can return when revisiting a particular result, while allowing different parts of the discussion to be read at different levels of detail. Readers interested primarily in the practical results may skip proofs and derivations, and in general \cref{sec:tri_lambda_tetr_g,sec:from_pol_to_gd,sec:proof_ps_boundary,sec:2_to_2_invitation,subsec:branch_selection} can be omitted on a first reading. 
In \cref{sec:gd} we introduce the language kinematic phase space are written in, Gram Determinants (GDs). We show how to interpret them geometrically and provide a dictionary with our most natural language for scatterings, energies and angles; we identify the scattering boundary as the place where GDs are 0. In \cref{sec:phase_space} we use these properties to characterize the $2\to3$ phase space. Once the phase space is known, we can compute cross section distributions on it. This is what we do in \cref{sec:xsec}. We conclude in \cref{sec:conclusion}.\\

\section{Gram Determinants}\label{sec:gd}
\subsection{Overview of Gram Determinants}\label{sec:gd_histo}
We can trace back the history of the use of the phase space properties for discovery back to Dalitz's work \cite{dalitz}. He wanted to characterise the $J^P$ of the $K^+$ from the $K^+ \to \pi^+ \pi^- \pi^+$ decay. Similar processes have only two degrees of freedom, that can be expressed in a Lorentz invariant form, for example, as invariant masses of two couples of final state particles. Basing on the population of a two dimensional plot of such variables, $J^P$ could be predicted. Dalitz was working in a non relativistic frame, but soon Fabri \cite{fabri} improved his work by implementing relativistic corrections.\\
The need for a mathematical grasp on the $1 \to n$ decay phase space was soon fulfilled by Byers and Yang: in \cite{byers_yang} they provide the phase space region and density via Gram Determinants. This is the foundational point of the theory of a generic phase space in a Lorentz invariant language, for this reason we summarize their work in \cref{sec:byers_yang}. However a $1\to n$ decay has $3n-7$ degrees of freedom and dealing with a $(3n-7)-$dimensional phase space density may be cumbersome. Therefore it is worth to specialize to some $n$ and study some phenomenologically relevant phase space distributions. This is the case of \cite{nyborg1965}, that considers $n=4$. Phase space distributions are provided for trivial matrix elements, emphasizing how symmetry can help us transform some distributions into some others. \cite{nyborg1965} also extend the validity of their results to $2 \to 4$ scatterings with special resonances.\\
Another specialization of \cite{byers_yang} is due to Morrow \cite{morrow}, studying the $2 \to 3$ scattering as a generalization of Kibble's treatment \cite{kibble}. He implicitly provides 2 dimensional distributions of the phase space (neglecting the matrix element). Kajantie and Lindblom \cite{2_to_3_finnici} start from Morrow's work and add to it the explicit analytic form of 1 and 2 dimensional phase space distributions.\\
Finally, what Byers and Yang express formally for the $1\to n$ generic decay is made explicit by Poon: in \cite{poon} the author derives all the determinantal relations which are necessary for expressing the $1 \to n$ phase space boundary.\\
The picture that emerges is that scattering phase spaces and their distributions are dense of information. If a scattering includes a BSM particle, model independent kinematic features can be used to recover the new particle mass. The first quantities that have been exploited in this sense are one dimensional kinematic features, namely edges and endpoints in invariant masses distributions \cite{Allanach:2000kt,Gjelsten:2004ki,Gjelsten:2005aw,Miller:2005zp,Matchev:2009iw,Burns:2009zi}. However \cite{agrawal} show that multi-dimensional features can be way more effective: the authors perform a likelihood analysis exploiting the integrable divergence on the phase space boundary of many-body processes and compare it to a quality of fit analysis of classical edges and endpoints of squared masses distributions. They obtain a great improvement on the latter in two directions: i) via multi-dimensional analysis one can determine not only NP mass differences but also the overall scale of masses; ii) the GD-based analysis is viable in case of low signal statistics too, which can likely be the case of BSM Physics.\\ 
Another direction for exploiting the dimension 4 Symmetric Gram Determinant \footnote{See \cref{subsec:definitions} for definition and properties.} $\Delta_4$ is found in \cite{Debnath:2016mwb}: they realize that the phase space boundary is both an accumulation surface for events thanks to the integrable divergence and a sharp limit beyond which signal events can not be found any more. To use this information it is necessary to study the fully derivative phase space, whose technical complication is overcome by the means of Voronoi Tessellations \cite{voronoi_book}. The $\Delta_4$ variable can itself be a Lorentz invariant function to leverage as a discovery variable at colliders, as proposed by \cite{Debnath:2018azt}.\\
Simplifying working assumptions in phase space literature until now has been neglecting masses and considering a constant matrix element. Such very powerful simplifications can capture fundamental aspects of the scattering but can not provide the full picture. If our goal is constraining feeble BSM signals, we must refine this pictures and take masses and matrix elements into account. For the sake of clarity we focus here on the simplest matrix element (pure propagators), more complicated cases being possible to tackle in this formalism but beyond the scope of this project. At the same time we could verify that there is a large ensemble of configurations for which pure propagators matrix elements yield a cross section with a negligible error with respect to the full matrix element, showing that our improvement over the state of the art is also sufficient.\\
We remark that the results of this work are not limited to specific topologies. In particular, in the past decades $2\to$ many scatterings were studied, also through Gram Determinants \cite{Gjelsten:2004ki,Burns:2009zi,Debnath:2018azt}, because decay chains initiated by supersymmtric particles have been of particular interest. In general, when dealing with decay chains, a solid strategy is considering subsystem of final state particles and use recursion relations \cite{ps_factoriz_original} on the phase space of these subsystems. \cite{Jing:2020tth} offer a graphical method to apply recursion with $n$ particles in $d$ dimensions. But New Physics is not necessarily hidden in decay chains: especially in this epoch of future high energy machines, $s-$channel like topologies are penalized with respect to VBF \cite{franceschini_2021_vbf} so it is imperative to find strategies to analytically describe the VBF too. It is true that in the extreme case of constant matrix element it can be shown that a decay chain framework is superior to a purely Lorentz invariant method if the aim is the computation of the fully integrated phase space: \cite{Hajer:2025gwr} explicitly show that decay chains naturally indicate invariant masses and angles between subsystems as the best suited set of variables to get to fully integrated cross sections because this set has the easiest phase space boundary. However not always can we completely neglect the matrix element and not always the fully integrated phase space is the aim of our computation, simply because that's not the most powerful discriminatory variable between signal and background: different propagators make different phase space regions orders of magnitudes more important than other and can guide us to the discovery of new particles. In this sense peaking cross section distributions can be way more important than the total phase space therefore it is mandatory to use the set of coordinates that enhances resonances the most. This work introduces two new topologies to the discussion, specifically realized here in \cref{fig:alp_chans}. This opens the possibility for many BSM models to be discovered thanks to Gram Determinants.

\subsection{Definitions}\label{subsec:definitions}
A recent reference for most of these properties is given in appendix C from \cite{tao_han_collider_pheno}. A more complete but older reference is appendix A from \cite{libro_finlandesi}.

\begin{definition}[Gram Matrix]

Consider a vector space and two sets of elements in this space, $\{p_1, \dots, p_n\}$ and $\{q_1, \dots, q_n\}$. For our purposes, let the $p_i, q_i$ be four-momenta and the vector space multiplication be the Minkowskian scalar product. Define the $n \times n$ Gram Matrix

\begin{equation}
\M{\{p_1, \dots, p_n\}}{\{q_1, \dots, q_n\}}=
\begin{pmatrix}
p_1 \cdot q_1 & \dots & p_1 \cdot q_n\\
\vdots & \vdots & \vdots\\
p_n \cdot q_1 & \dots & p_n \cdot q_n
\end{pmatrix}.
\end{equation}
\end{definition}

\begin{definition}[Gram Determinants (GDs)]\label{def:gd}

A GD is the determinant of a Gram Matrix:

\begin{equation}
\G{\{p_1, \dots, p_n\}}{\{q_1, \dots, q_n\}}=\det \M{\{p_1, \dots, p_n\}}{\{q_1, \dots, q_n\}}
\end{equation}
\end{definition}

\begin{definition}[Symmetric Gram Determinant.]
~\\
The symmetric Gram Determinant (sGD) is a GD whose two entries are the same.
\begin{equation}
\Delta_n(\{p_1, \dots, p_n\})=\G{\{p_1, \dots, p_n\}}{\{p_1, \dots, p_n\}}.
\end{equation}
Call $n$ the dimension of $\Delta_n$.
\end{definition}

\begin{lemma}[Invariance under argument ordering]\label{lemma_in_string_permutations}

The order of the two GD arguments is irrelevant because of the invariance of the determinant of a matrix under exchange of its rows or columns.

\begin{align}
&\G{\{p_1, \dots, p_i, p_{i+1}, \dots, p_n\}}{\{q_1, \dots, q_n\}}=\G{\{p_1, \dots, p_{i+1}, p_i \dots, p_n\}}{\{q_1, \dots, q_n\}},\\
&\G{\{p_1, \dots, p_n\}}{\{q_1, \dots, q_i, q_{i+1}, \dots, q_n\}}=\G{\{p_1, \dots, p_n\}}{\{q_1, \dots, q_{i+1}, q_i, \dots, q_n\}}.
\end{align}

\end{lemma}

\begin{lemma}[Invariance on the order of the arguments]\label{lemma_string_permutation}

The order of the arguments of a GD is irrelevant because a matrix determinant is invariant under matrix transposition.

\begin{equation}
\G{\{p_i\}_{i=1}^n}{\{q_i\}_{i=1}^n}=\G{\{q_i\}_{i=1}^n}{\{p_i\}_{i=1}^n}.
\end{equation}

\end{lemma}

\begin{lemma}[One momentum deformation]

If any of the elements of any GD argument is multiplied by a scalar $\lambda$ of the vector space, the whole GD is rescaled by the same scalar:

\begin{align}
&\G{\{p_1, \dots, \lambda p_i, \dots, p_n\}}{\{q_1, \dots, q_n\}}= \lambda \G{\{p_1, \dots,  p_i, \dots, p_n\}}{\{q_1, \dots, q_n\}} \label{lambda_row},\\
&\G{\{p_1,  \dots, p_n\}}{\{q_1, \dots, \lambda q_i, \dots, q_n\}}= \lambda \G{\{p_1, \dots,  p_i, \dots, p_n\}}{\{q_1, \dots, q_n\}} \label{lambda_column},
\end{align}

\noindent
because $G$ is the $\det$ of a matrix whose whole row (\cref{lambda_row}) or column (\cref{lambda_column}) is rescaled.

\end{lemma}

\begin{lemma}[Invariance under linear combination of argument elements]\label{lemma:sum_one_mom}

A GD is left invariant if any of its arguments is replaced by the same set in which one entry is traded for a linear superposition of its elements:

\begin{align}
&\begin{aligned}
\G{\{p_1, \dots, p_i, \dots, p_n\}}{\{q_1, \dots, q_j, \dots, q_n\}}&=\G{\{p_1, \dots, p_i', \dots, p_n\}}{\{q_1, \dots, q_j, \dots, q_n\}}\\
&=\G{\{p_1, \dots, p_n\}}{\{q_1, \dots, q_j', \dots q_n\}}
\end{aligned},\\
&p_i'=\sum_{a=1}^n c_a p_a, \qquad q_j'=\sum_{a=1}^n g_a q_a, \qquad \vec{c}, \vec{g} \neq \vec{0}
\end{align}
This is due to the invariance of the determinant of a matrix under linear superposition of its rows or columns.
\end{lemma}

\begin{lemma}[Principal Gram Determinant (PGD) of a scattering]
In an $n \to N-n$ process the PGD is a symmetric GD computed onto whatever subset of four-momenta of dimension $N-1$. This is due to \cref{lemma:sum_one_mom}.
\end{lemma}

\subsection{Geometrical meaning of Gram Determinants}\label{sec:tri_lambda_tetr_g}

We might be used to think of the kinematic phase space as the object in a cross section integral enforcing the conservation of four-momentum, $\delta^{(4)}(p_\text{tot})$. Like this, this Dirac $\delta$ distribution looks rather abstract, while it is more useful to think of the kinematic phase space as a convex volume in the space of final state particles energies and angles (or equivalently in the space of possible four-momenta scalar products). In this way, even without computing any of them, we know that kinematic quantities have upper and lower boundaries and that the boundary on one variable depends on other variables. If the volume is finite, a cross section can not diverge because of the phase space, it can only diverge if the kinematic configurations leading to divergence fall within the phase space. We can even estimate that a scattering whose poles are closer to the phase space volume will be more important than those with further poles. Gram Determinants are exactly the tool that can help us see the phase space as a volume. The first example of this is the triangular $\lambda$, corresponding both to a $\Delta_2$ and to the volume of a triangle. In three dimensions we can define an analogous function, the tetrahedral or kinematic ${\cal G}$, corresponding to $\Delta_3$. The reason for this GD-volume correspondence is that matrix determinants can be associated with the volume of a hyper-rectangle (\cref{parallelepiped_remark}). More generally, we can make GDs correspond to Cayley determinants: simplistically a Cayley determinant is a function that computes the volume of, so to say, $n$ dimensional triangles (simplexes). 

\begin{remark}[Determinants of an $n \times n$ matrix $M$ and volume of a parallelepiped in $\mathbb{R}^n$]\label{parallelepiped_remark}
Consider an $n \times n$ diagonal matrix $M_D$. You can see its eigenvalues as the sides of a hyper-rectangle in $n$ dimensions and its determinant as its volume. If you transform 
\begin{equation}
M_D \to M =A \cdot M_D \cdot A^{-1},
\end{equation}

\noindent
with $A$ an invertible $n \times n$ matrix, the determinant is unchanged. However the columns of $M_D$ were orthogonal vectors while those of $M$ are not. This means that the corresponding parallelepiped is no longer rectangle. Still, the determinant of an $n \times n$ matrix is the volume of a hyper-rectangle with one vertex in the origin and all other vertices pointed by the matrix columns.
\end{remark}

\noindent
But we can do more than just rectangles:

\begin{definition}[Cayley determinants]\label{cayley_def}
Cayley determinants are a way of expressing the area of a simplex\footnote{A $k-$simplex is a $k-$dimensional polytope, i.e. a geometric object with flat faces. In other words, given $k+1$ vertices, the $k-$simplex is their smallest convex envelope.}. Suppose to have $n+1$ points in $\mathbb{R}^k$ with $k \geq n$. They define an $n-$dimensional simplex. Call $V_n$ its volume and $d_{ij}$ the distance among the $i^\text{th}$ and the $j^\text{th}$ point. Build the following matrix $M_n$:

\begin{equation}
M_{n-1}=
\begin{pmatrix}
0 & 1 & 1 & \dots &1\\
1 & 0 & d_{12} & \dots & d_{1n}\\
& & \vdots &\\
1 & d_{1n} & d_{2n} & \dots & 0
\end{pmatrix}.\label{eq:cayley_matrix}
\end{equation}
Then

\begin{equation}
V_n^2=(-1)^{n+1} \dfrac{\det M_n}{\Gamma(n+1)^2 2^n}.
\label{n_dim_cayley_volume}
\end{equation}
\end{definition}

\noindent
Gram and Cayley determinants can be equivalently used, we only need a rule for expressing the $d_{ij}$ in \cref{eq:cayley_matrix}. For this purpose let me define:

\begin{definition}[$m-$particle invariant masses]\label{inv_mass_def}
\begin{equation}
s_{i j k \dots}:=(p_i+p_j+p_k + \dots)^2.
\end{equation}
Ingoing and outgoing particles have opposite signs.
\end{definition}

\noindent
From this we derive the following lemma

\begin{lemma}[Relation between the $n-$dimensional Cayley volume and the $\Delta_n$]\label{lemma:gram_cayley}
There is a 1 to 1 correspondence between $\Delta_n$ and the Cayley determinant built as in \cref{cayley_def} with

\begin{equation}
\Delta_n(p_1, \dots, p_n)=\Gamma(n+1)^2 V_n^2({d}_{ij})
\end{equation}

\noindent
if sides are chosen like
\begin{equation}
d_{ij}=s_{i, (i+1), (i+2), \dots, (j-1)}= \left(\sum\limits_{k=i}^{j-1} p_k\right)^2, i \in [1,n], j \in [2,n+1].\label{eq:choose_sides}
\end{equation}

\noindent
Signs are as in \cref{inv_mass_def}.
\end{lemma}

\noindent
We will not prove \cref{lemma:gram_cayley} but one can easily check it for $n=1$ (trivially true) and for $n=2$. On one hand

\begin{equation}
\Delta_2=
\begin{vmatrix}
m_1^2 & p_1 \cdot p_2\\
p_1 \cdot p_2 & m_2^2
\end{vmatrix}
= (m_1 m_2)^2- (p_1 \cdot p_2)^2.\label{eq:explicit_d2}
\end{equation}

\noindent
On the other hand, using \cref{eq:choose_sides},

\begin{equation}
V_2^2=-\dfrac{1}{16}
\begin{vmatrix}
0 & 1 & 1 & 1\\
1 & 0 & m_1^2 & s_{12}\\
1 & m_1^2 & 0 & m_2^2\\
1 & s_{12} & m_2^2 & 0
\end{vmatrix}=\dfrac{(m_1 m_2)^2- (p_1 \cdot p_2)^2}{4},
\end{equation}

\noindent
which satisfies \cref{lemma:gram_cayley}.\\


The first non trivial simplex is the triangle. Correspondingly we define an ubiquitous function of particle physics:

\begin{definition}[Triangular $\lambda$]\label{def:tri_lambda}
\begin{equation}
\lambda(x,y,z)=(x-y-z)^2-4 y z,\label{eq:tri_lambda}
\end{equation}
\Cref{eq:tri_lambda} is fully symmetric under all $x, y, z$ permutations.
\end{definition}

\noindent
The reason for it to be called triangular can be found in the following theorem:

\begin{theorem}[The triangular $\lambda$ is the area of a triangle]
From the Hero theorem $\dfrac{\sqrt{-\lambda(x^2,y^2,z^2)}}{4}$ is the area of the triangle of sides $x,y,z$.
\end{theorem}

\begin{corollary}[$\Delta_2$ is the area of a triangle]\label{corol:d2_area_triangle}
\begin{align}
&4 \Delta_2(p_1,p_2)^2=-\lambda(x,y,z),\\
&x:=m_1^4+(p_1\cdot p_2)^2,\\
&y:=(p_1\cdot p_2)^2+m_2^4,\\
&z:=(m_1^2-(p_1\cdot p_2))^2+(m_2^2-(p_1\cdot p_2))^2.
\end{align}
\end{corollary}

\noindent
Following the same philosophy, the next simplex is the tetrahedron:

\begin{definition}[Kinematic ${\cal G}$]\label{def_kin_g}
\begin{equation}
{\cal G}(x,y,z,u,v,w):=
\begin{vmatrix}
0 & 1 & 1 & 1 & 1 \\
1 & 0 & u & y & v \\
1 & u & 0 & w & x \\
1 & y & w & 0 & z \\
1 & v & x & z & 0 \\
\end{vmatrix}.
\end{equation}
\end{definition}

\begin{theorem}[The tetrahedral ${\cal G}$ and $\Delta_3$ are the area of a tetrahedron]\label{theo:d3_thetrahedron}
There is a 1 to 1 correspondence between $\Delta_3$ and ${\cal G}$:

\begin{equation}
\Delta_3(p_1,p_2,p_3)=-\dfrac{1}{2} {\cal G}((p_1+p_2)^2,(p_1-p_3)^2,(p_1+p_2-p_3)^2,m_1^2,m_2^2,m_3^2).
\end{equation}

\noindent
Consequently $\Delta_3$ is the area of the tetrahedron whose vertices are

\begin{align}
&v_1=(0,0,0),\\
&v_2=(0,0,u),\\
&v_3=y(\sin \theta_y,0,\cos \theta_y),\\
&v_4=v(\sin \theta_v \cos \phi, \sin \theta_v \sin \phi, \cos \theta_v),
\end{align}

\noindent
with $(\theta_y,\theta_v, \phi)$ solutions of

\begin{align}
& \norm{v_2-v_3}=w,\\
& \norm{v_2-v_4}=x,\\
&\norm{v_3-v_4}=z.
\end{align}

\end{theorem}

\subsection{From polar coordinates to Lorentz invariant variables}\label{sec:from_pol_to_gd}

\begin{lemma}[Transformation of polar coordinates to sGDs]\label{dictionary_lemma}

Let $p_i$ have mass $m_i$, energy $E_i$, polar angle $\theta_i$ and azimuthal angle $\phi_i$

\begin{equation}
\begin{aligned}
&p_i=(E_i, \vec{p}_i)=(E_i, |\vec{p}_i| \sin \theta_i \cos \phi_i, |\vec{p}_i| \sin \theta_i \sin \phi_i, |\vec{p}_i| \cos \theta_i),\\
&|\vec{p}_i|^2=E_i^2-m_i^2.
\end{aligned}
\end{equation}
There is a non bijective relation among the $(m_i,E_i, \theta_i, \phi_i)$ and the $\Delta_{1},\Delta_{2},\Delta_{3},\Delta_{4}$ defined on the $p_i$. Consider a set of four-momenta $\{p_i\}_{i=1}^n$ in 4 dimensions and go to $p_0$ rest frame. Then

\begin{align}
& m_i = \sqrt{\Delta_1(p_i)},\\
& |\vec{p}_i|= \dfrac{\sqrt{-\Delta_2(p_0, p_i)}}{m_0},\\
& |\sin \theta_{ij}|= m_0 \sqrt{\dfrac{\Delta_3(p_0,p_i,p_j)}{\Delta_2(p_0, p_i)\Delta_2(p_0, p_j)}} \label{s_t_ij},\\
& |\sin \phi_{(ij);(ik)}|=\left[\dfrac{\Delta_4(p_0,p_i,p_j,p_k) \Delta_2(p_0,p_i)}{\Delta_3(p_0,p_i,p_j)\Delta_3(p_0,p_i,p_k)}\right]^{1/2} \label{s_f_ij_ik}.
\end{align}

\noindent
This can be seen either from direct computation or found in appendix A of \cite{libro_finlandesi}, for example. In \cref{s_t_ij}, $\theta_{ij}$ is the angle between $p_i$ and $p_j$. For example, if $\vec{p}_i$ is along the $z$ axis, $\theta_{ij}$ is the polar angle of $\vec{p}_j$. In \cref{s_f_ij_ik}, $\phi_{(ij);(ik)}$ is the angle between two planes spanned by $(\vec{p}_i,\vec{p}_j)$ and $(\vec{p}_i,\vec{p}_k)$ respectively. For example, if $\vec{p}_i$ is along the $z$ axis and $\vec{p}_j$ is along the $x$ axis, $\phi_{(ij);(ik)}$ is the azimuthal angle of $\vec{p}_k$.
\end{lemma}

\begin{theorem}[Signs and bounds]\label{signs_bounds_theorem}
Given the \cref{dictionary_lemma} and the fact that

\begin{equation}
m_i,|\vec{p}_i|, |\sin \theta_{ij}|,|\sin \phi_{(ij);(ik)}| \in \mathbb{R}^+ \ \forall \ i,j,
\end{equation}

\noindent
the following are true:

\begin{subequations}\label{signs_and_bounds}
\begin{align}
& \Delta_1(p_i) \geq 0,\\
& \Delta_2(p_0, p_i)\leq 0,\\
& \Delta_3(p_0,p_i,p_j) \geq 0,\\
& \Delta_4(p_0,p_i,p_j,p_k) \leq 0.
\end{align}
\end{subequations}
\end{theorem}

\subsection{Null Gram determinants as the boundary of the physical phase space}\label{sec:proof_ps_boundary}

In the previous section we showed how to perform a change of variables from polar coordinates to Lorentz invariant quantities. But how do we express the phase space boundary? What is the cross section integration domain? Starting from the easiest 2 bodies case to the most generic configuration, we will show that the phase space boundary is the locus of the nodes of the Principal Gram Determinant of the scattering. This is why GDs are the most natural way of expressing a scattering phase space.

\subsubsection*{2 bodies}\label{subsec:2_body}
Consider the four-momenta $p_a$ and $p_b$ and go to the rest frame of the former; call $\vec{\beta}$ the velocity of $p_b$ in this frame and $\gamma$ its associated Lorentz factor. Then we specialize \cref{eq:explicit_d2} to

\begin{equation}
\Delta_2(p_a,p_b)=-m_a m_b \gamma^2 \beta^2=0 \text{ if } \beta=0.
\end{equation}

\noindent
This means $\vec{p}_a,\vec{p}_b$ define a 0-dimensional simplex (a \textbf{point}).\\

\subsubsection*{3 bodies}\label{subsec:3_body}
Consider now the process $p_1 \to p_2 + p_3$. Three-momenta are arranged along the sides of a triangle because $\vec{p}_1=\vec{p}_2+\vec{p}_3$. Go to the reference frame in which one momentum is at rest, for example $p_1$ (in this problem indices are irrelevant, $1 \leftrightarrow 2 \leftrightarrow 3$). This point must belong to the boundary of the physical region because $\vec{0}$ is the minimum allowed to the $\vec{p}_i$ . Then $|\vec{p}_2+ \vec{p}_3|=0$ too, i.e. the velocities $\vec{\beta}_2$ and $\vec{\beta}_3$ are collinear and the triangle degenerated into a segment. We can state the following theorem:

\begin{theorem}[Degenerate triangle and maximal energies]\label{theo:deg_triang}
In the process $p_1 \to p_2 + p_3$, $|\vec{p}_2|$ is maximal when $|\vec{p}_1|=0$.
\end{theorem}

\begin{figure}[!h]
\centering
\includegraphics[width=\textwidth]{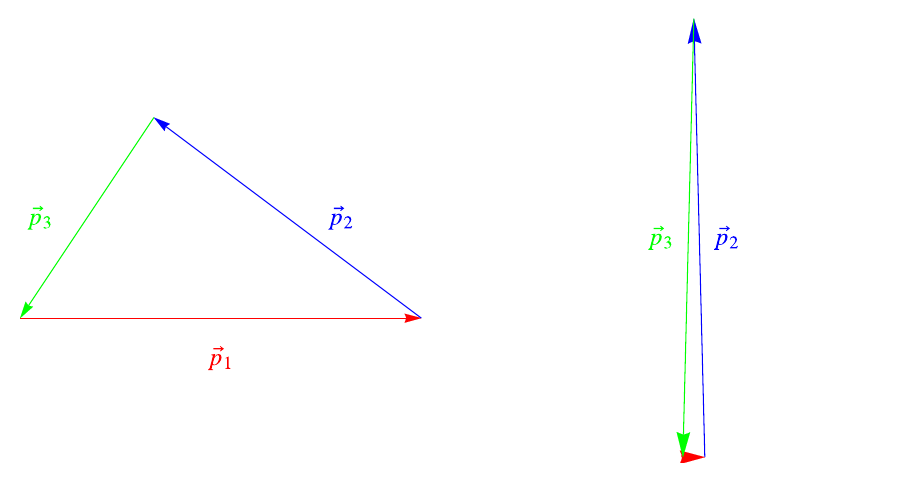}
\caption{Graphic representation of the three-momenta $\vec{p}_{1,2,3}$ satisfying $\vec{p}_1-\vec{p}_2-\vec{p}_3=\vec{0}$. Left: generic frame; right: $p_1$ rest frame (a small $\vec{p}_1$ magnitude is left for visualization purpose).}
\label{fig:kin_triangles}
\end{figure}

\noindent
Let us prove \cref{theo:deg_triang}. In $p_1$ rest frame (that we will call CoM and indicate with a $^*$) $|\vec{p}^*_1|=0$ and $|\vec{p}^*_2|$ and $|\vec{p}^*_3|$ are

\begin{align}
&|\vec{p}^*_2|=|\vec{p}^*_3|=\dfrac{\sqrt{\lambda_0}}{2 m_1}=:p,\\
&\lambda_0:=\lambda(m_1^2,m_2^2,m_3^2),
\end{align}
see \cref{fig:kin_triangles}. We want to show that if we perform a boost of Lorentz parameter $\beta$ to a new rest frame, the modulus of $\vec{p}_2$ will be the biggest if the boost has an angle $\theta=0$ with respect to $\vec{p}_2$, i.e. if the boost is parallel to $\vec{p}_2$, regardless from $\beta$. The new $p_2$ is a function of $\beta$ and $\theta$:

\begin{equation}
2 m_1 \sqrt{1-\beta^2} p_2=\left(\Sigma+\beta \cos \theta \sqrt{\lambda_0},\sin \theta \sqrt{\lambda_0}\sqrt{1-\beta^2}, 0, \beta \Sigma+ \cos \theta \sqrt{\lambda_0}\right)
\end{equation}
with $\Sigma:=m_1^2+m_2^2-m_3^2$. The derivative of $|\vec{p}_2|$with respect to $\theta$ is

\begin{equation}
\dv{|\vec{p}|^2}{\theta}= -\dfrac{p \ \beta \ \sin \theta}{m_1 (1-\beta^2)} \left[m_1^2\left(1+2 \beta \cos \theta \dfrac{p}{m_1}\right)+m_2^2-m_3^2\right] \leq 0 \ \forall \theta \in [0,\pi], \ \forall \beta \in [0,1],\label{eq:p_dv_theta}
\end{equation}


\noindent
because $m_1>m_2+m_3$. \Cref{eq:p_dv_theta} vanishes onto $\beta=0$, corresponding to making no boost, when the square bracket is 0, corresponding to two unacceptable solutions on which $p_2^0=0<m_2$, and onto $\theta=0$. This proves our statement. Let us remark that this demonstration only depends on the fact that $\vec{p}_{1,2,3}$ form a triangle.

\noindent
The corresponding GD in $p_1$ rest frame is

\begin{equation}
\Delta_3(p_1,p_2,p_3)=(m_1 |\vec{p}_2| |\vec{p}_3| \sin \theta_{23})^2,
\label{polar_delta_3}
\end{equation}
where $\theta_{23}$ is the angle between $\vec{p_2}$ and $\vec{p_3}$. It evaluates to 0 if $\sin \theta_{23}=0$, corresponding to \textbf{aligned} velocities.\\

\subsubsection*{4 bodies}\label{subsec:4_body}
\begin{figure}[!h]
\centering
\includegraphics[width=\textwidth]{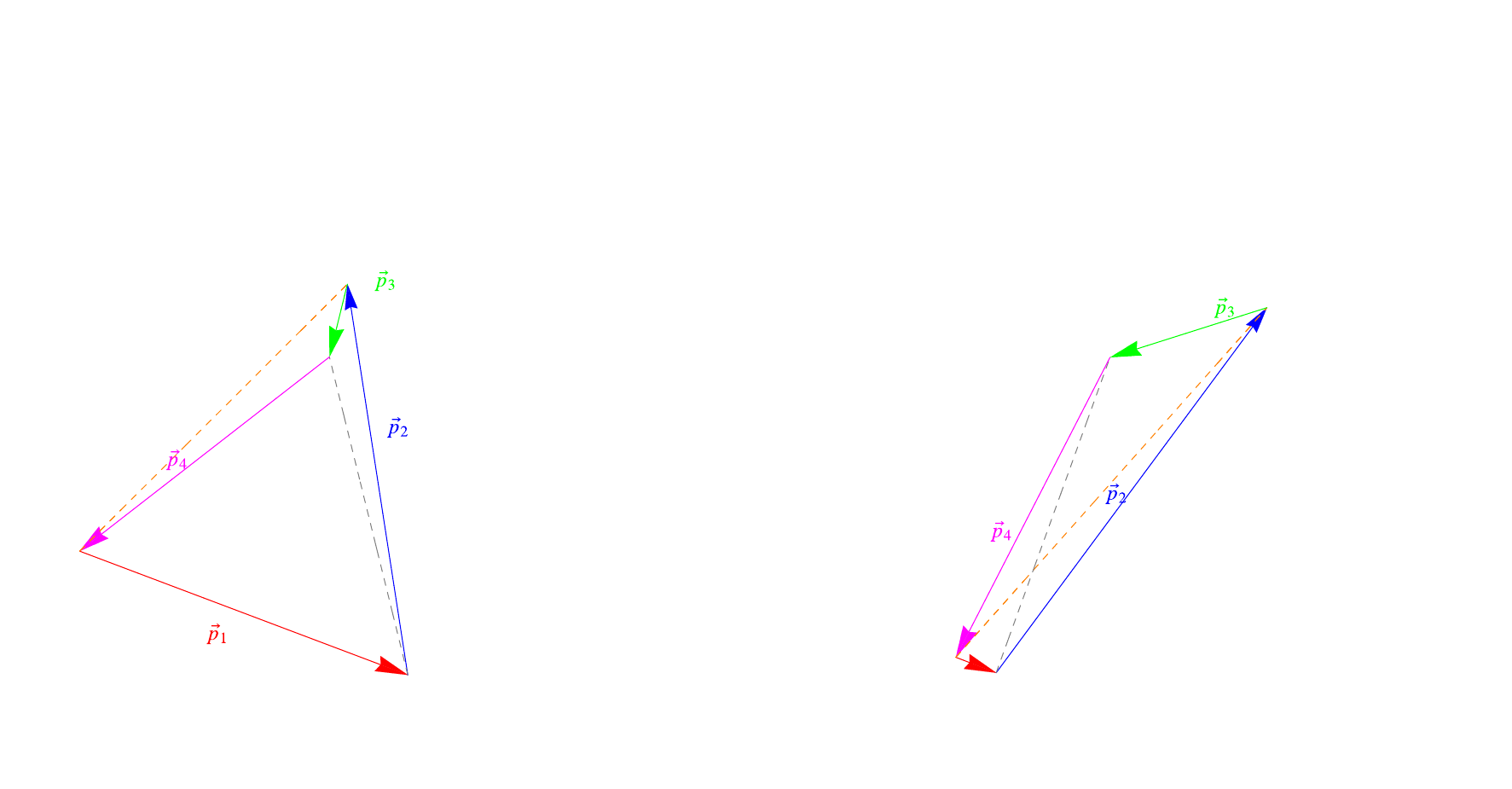}
\caption{Graphic representation of the three-momenta $\vec{p}_{1,2,3,4}$ satisfying $\vec{p}_1+\vec{p}_2=\vec{p}_3+\vec{p}_4$. Left: generic frame; right: $p_1$ rest frame. A small $\vec{p}_1$ magnitude is left for visualisation purpose. The dashed sides do not correspond to any four-momentum: the orange one corresponds to the sum  $\vec{p}_1+\vec{p}_2$ (or equivalently $\vec{p}_3+\vec{p}_4$), the grey one simply closes the tetrahedron.}
\label{fig:kin_tetras}
\end{figure}

Consider now the process $p_1 + p_2 \to p_3 + p_4$. Three-momenta are arranged along the sides of a tetrahedron, see \cref{fig:kin_tetras}. Call $\tau_{ij}$ the triangular face containing the three-momenta $\vec{p}_{i,j}$, the third follows.\\
In $p_1$ rest frame the tetrahedron degenerates into a triangle, $\tau_{23}$ collapsed onto $\tau_{34}$, $\tau_{14}$ becomes a segment. All other three-momenta are maximized with respect to $\vec{p}_1$, as from \cref{theo:deg_triang}. $\vec{\beta}_2, ~ \vec{\beta}_3$ and $\vec{\beta}_4$ are \textbf{coplanar} (they belong to the $\tau_{23} \equiv \tau_{34}$ plane).

Correspondingly, let us go in $p_1$ rest frame and assume none of the four-momenta is aligned with the other; as above, name $\theta_{ab}$ the angle among $\vec{p}_a$ and $\vec{p}_b$; define $\phi$ the angle among the planes generated by the couples $(\vec{p_2},\vec{p_3})$ and $(\vec{p_2},\vec{p_4})$, then 

\begin{equation}
\Delta_4(p_1,p_2,p_3,p_4)=-(m_1 |\vec{p}_2| |\vec{p}_3| |\vec{p}_4| \sin \theta_{23} \sin \theta_{24} \sin \phi)^2= 0 \text{ if } \phi=0. 
\label{polar_delta_4}
\end{equation}

\noindent
This is the definition of \textbf{coplanarity} of $\vec{p}_2,\vec{p}_3,\vec{p}_4$.

\subsubsection*{$n$ bodies}\label{subsec:generic_body}

We now generalize what we saw for 2, 3 and 4 particles for a generic process with $n$ initial and $m$ final state particles. This proof follows the one in section 3.2.1 from \cite{libro_cinematica}. The idea is that we can pick whatever subset of particles and go to the subset rest frame. This is an extremal configuration, in the sense that it belongs to the phase space boundary. Exploiting the fact that the system is at rest, we will be able to write down a system of equations involving the particles four-momenta. The system is solved non trivially when the Principal Gram Determinant of the scattering vanishes. 
\begin{theorem}[Symmetric Gram Determinants are null when there is a linear dependence among their arguments.]
Consider a scattering with $n$ initial state particles $\{p_i\}_{i=1}^n$ and $m$ final state particles $\{p_j\}_{j=n+1}^N, n+m=N$:

\begin{equation}
p_1 + \dots + p_n \to p_{n+1}+ \dots + p_N.
\end{equation}

\noindent
Four-momentum conservation is written as

\begin{equation}
\sum\limits_{i=1}^n p_i= \sum\limits_{j=n+1}^N p_j.
\end{equation}

\noindent
To write simpler sums let us write an equivalent process with all initial state particles

\begin{equation}
q_1 + \dots + q_N \to 0,
\end{equation}

\noindent
where

\begin{equation}
q_i:=
\begin{cases}
p_i & \text{ if } 1\leq i \leq n,\\
-p_i & \text{ if } n< i \leq N.\\
\end{cases}
\end{equation}

\noindent
Among the $q_i$, pick a group of $k$ four-momenta and say they make up the system ${\cal S}$; go to its rest frame, where the four-momentum of ${\cal S}$ is $q_{{\cal S}}=(M_{{\cal S}},\vec{0})$. Then

\begin{equation}
\sum\limits_{i \notin {\cal S}} q_i+\sum\limits_{i \in {\cal S}} q_i= \sum\limits_{i \notin {\cal S}} q_i+ q_{{\cal S}}=(0,\vec{0}), \label{sys_sum}
\end{equation}

\noindent
which is specifically true for the time component

\begin{equation}
\sum\limits_{i \notin {\cal S}} q_i^0=-M_{{\cal S}}.
\end{equation}

\noindent
Let's rewrite the unity as

\begin{equation}
1 = -\dfrac{1}{M_{{\cal S}}} \sum\limits_{i \notin {\cal S}} q_i^0, \label{one_mass}
\end{equation}

\noindent
and use it in \cref{sys_sum}:

\begin{equation}
\sum\limits_{i \notin {\cal S}} q_i= \left(\dfrac{1}{M_{{\cal S}}} \sum\limits_{i \notin {\cal S}} q_i^0\right) q_{{\cal S}}.
\label{lin_dep}
\end{equation}

\noindent
We can rewrite \cref{lin_dep} as a liner combination of all the $q_i$ by using again that $\sum\limits_{i \in {\cal S}} q_i= q_{{\cal S}}$:

\begin{align}
&\sum\limits_{i=1}^N\lambda_i q_i=0,\\
&\lambda_i = 
\begin{cases}
1 & \text{ if } i \notin {\cal S},\\
-\dfrac{1}{M_{{\cal S}}} \sum\limits_{i \notin {\cal S}} q_i^0 & \text{ if } i \in {\cal S}.
\end{cases}
\end{align}

\noindent
Now if $\sum\limits_{i=1}^N\lambda_i q_i=0$, then also $q_i \cdot \sum\limits_{i=1}^N\lambda_i q_i=0$ for any $i$. We can collect all these equations in the system:

\begin{equation}
\begin{cases}
& q_1 \cdot \sum\limits_{i=1}^N\lambda_i q_i=0,\\
& \dots ,\\
& q_{N-1} \cdot \sum\limits_{i=1}^N\lambda_i q_i=0.\\
\end{cases}
\end{equation}

\noindent
Notice that the $N^\text{th}$ equation is missing because the equation for $q_N$ can be recovered from four-momentum conservation. The determinant of the system for the unknowns $\lambda_i$ is exactly the Principal Gram Determinant of the scattering. The solution for the system in non trivial if the PGD is 0.\\
This concludes our demonstration of the fact that null Gram Determinants define the boundary of the phase space region.
\end{theorem}

\section{Phase Spaces}\label{sec:phase_space}

In this section we recap a Lorentz invariant treatment of the $2\to2, 1 \to3$ and $2\to3$ phase spaces, mostly basing on \cite{libro_finlandesi,libro_cinematica}.

\subsection{$2 \to 2$ scattering}\label{sec:2_to_2_invitation}

In this section we work out \cref{signs_and_bounds} for a $2\to2$ scattering $p_1+p_2 \to p_3 + p_4$. It is standard to define the Mandelstam variables as

\begin{equation}
\begin{aligned}
&s=(p_1+p_2)^2,\\
&t=(p_2-p_4)^2,\\
&u=(p_1-p_4)^2.
\end{aligned}\label{def:s_t_u}
\end{equation}

\noindent
From four-momentum conservation it follows that

\begin{equation}
s+t+u=m_1^2+m_2^2+m_3^2+m_4^2=:k.\label{eq:k_def}
\end{equation}

\noindent
In each row of \cref{tab:2_to_2_d2_constr} we list the effect of imposing $\Delta_2(p_i,p_j)\leq 0$.

\begin{table}[!h]
\begin{tabular}{c|c}
$(i,j)$ & Bound \\ 
\hline
$(1,2) \equiv (3,4)$ & $s \leq (m_1-m_2)^2 \cup s \geq (m_1+m_2)^2$ \\ 
(1,3) & $t \leq (m_1-m_3)^2 \cup t \geq (m_1+m_3)^2$ \\ 
(1,4) & $u \leq (m_1-m_4)^2 \cup u \geq (m_1+m_4)^2$ \\ 
(2,3) & $u \leq (m_2-m_3)^2 \cup u \geq (m_2+m_3)^2$ \\ 
(2,4) & $t \leq (m_2-m_4)^2 \cup t \geq (m_2+m_4)^2$ \\ 
$(1+2,3) \equiv (3+4,3) \equiv (3,4) \equiv (1+2,4)$ & $s \leq (m_3-m_4)^2 \cup s \geq (m_3+m_4)^2$ \\ 
\end{tabular}
\caption{On the left column the couple of four-momenta entering in $\Delta_2(p_i,p_j)$. On the right column the bounds on $s, t$ or $u$ coming from $\Delta_2(p_i,p_j)\leq 0$.}
\label{tab:2_to_2_d2_constr}
\end{table}

\bigskip

\begin{figure}[H]
\centering
\includegraphics[width=0.65\linewidth]{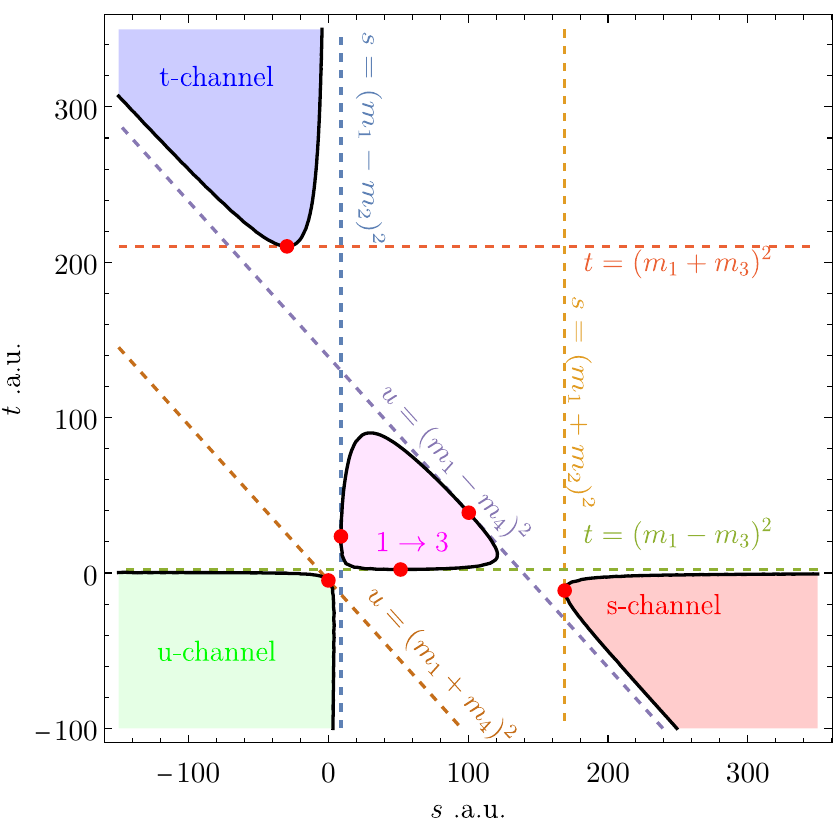}
\includegraphics[width=0.65\linewidth]{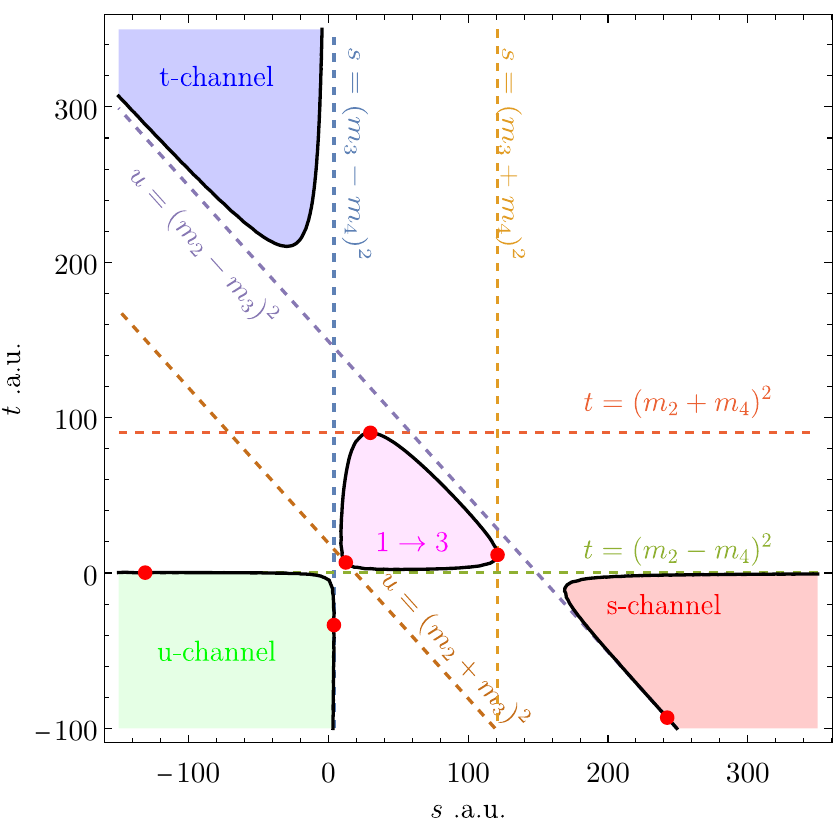}
\caption{Example of Kibble curve for the mass choice $m_1=8, m_2=5,m_3=6.5,m_4=4.5$ a.u. For clarity we split the asymptotes in two pictures. In both panels the red, blue and green area and the black curve correspond to $t-, s-$ and $u-$channel and to the Kibble curve respectively. Red dots are tangency points.}
\label{fig:Kibble}
\end{figure}

The Principal Gram Determinant for a $2 \to 2$ scattering is $\Delta_3$, equivalently ${\cal G}(s, t, m_4^2,m_1^2,m_2^2,m_3^2)$, or simply ${\cal G}$. ${\cal G}\leq0$ defines the scattering phase space, ${\cal G}=0$ is its boundary. It is also called Kibble curve \cite{kibble}

\begin{equation}
-2 {\cal G}= s t u- (\alpha s+ \beta t + \gamma u)=0, \label{Kibble_def}
\end{equation}

\noindent
with

\begin{align}
& k \alpha := (m_1^2 m_2^2- m_3^2 m_4^2)(m_1^2+m_2^2-m_3^2-m_4^2), \label{alpha_def}\\
& k \beta := (m_1^2 m_3^2- m_2^2 m_4^2)(m_1^2+m_3^2-m_2^2-m_4^2), \label{beta_def}\\
& k \gamma := (m_1^2 m_4^2- m_2^2 m_3^2)(m_1^2+m_4^2-m_2^2-m_3^2). \label{gamma_def}
\end{align}

\noindent
where $k$ is defined in \cref{eq:k_def}. The Kibble curve and its tangents are represented in \cref{fig:Kibble}. The area identified by ${\cal G}\geq0$ in the $(s,t)$ plane is a disconnected union of four convex regions. Three of them correspond to the $2 \to 2$ scattering and one to the $1\to 3$ decay. The latter is finite, while the former are only limited by two tangents. Each of them correspond to the $s,t$ or $u$ channel. The $s$ region is located in the $s\geq(m_1+m_2)^2, \ t \leq (m_1-m_3)^2$ portion of the $(s,t)$ plane; the horizontal line $t=0$ is an asymptote to this region in the $s\to\infty$ limit. The $t$ region has $t\geq(m_1+m_3)^2, \ s\leq(m_1-m_2)^2$ and the vertical line $s=0$ is an asymptote for the $t\to \infty$ limit. However the Kibble curve intersects the lines $s=0, t=0, u=0$ before approaching them at infinity. Let us derive the actual envelope. 

\noindent
The system

\begin{equation}
\begin{cases}
&{\cal G}=0\\
&s+t+u=k
\end{cases}
\end{equation}
(i.e. \cref{Kibble_def,eq:k_def}) is solved, for example, by 

\begin{equation}
\begin{cases}
&t=t(s)\\
&u=k-s-t(s)
\end{cases}
\end{equation}
$t(s)$ is the Kibble curve. To find its tangents we can explicitly find $t(s)$ and solve $\dv{t(s)}{s}=0$. An easier way is recovering $\dv{t(s)}{s}$ from $\dv{{\cal G}}{s}$:

\begin{equation}
\dv{{\cal G}}{s}=\pdv{{\cal G}}{s}+\pdv{{\cal G}}{t}\dv{t}{s}+\pdv{{\cal G}}{u}\dv{u}{s}=0 \label{eq:kibble_der}.
\end{equation}
Explicitly

\begin{align}
&\pdv{{\cal G}}{s}= t u - \alpha,\\
&\pdv{{\cal G}}{t} = s u - \beta,\\
&\pdv{{\cal G}}{u}= s t - \gamma,\\
&\dv{u}{s}=-1-\dv{t}{s}.\\
\end{align}

\noindent
By plugging the latter into \cref{eq:kibble_der}

\begin{equation}
\dv{{\cal G}}{s}=t u - \alpha+ (s u - \beta)\dv{t}{s}-(s t - \gamma)\left(1+\dv{t}{s}\right)=0,
\end{equation}

\noindent
which is solved by

\begin{equation}
\dv{t}{s}=-\dfrac{s t - \gamma- tu +\alpha}{s t - \gamma- su +\beta}\label{eq:dt_ds}.
\end{equation}

\noindent
\Cref{eq:dt_ds} is 0 while satisfying ${\cal G}=0, \ s+ t+u=k$ onto

\begin{equation}
t=(m_1 \pm m_3)^2,(m_2 \pm m_4)^2.
\end{equation}

\noindent
All other tangents can be found with an analogous procedure:

\begin{table}[htbp]
\centering
\begin{tabular}{c|cc}
Tangent parallel to plane & \multicolumn{ 2}{c}{Solutions} \\ 
\hline
$s=0$ & $s=(m_1 \pm m_2)^2,$ & $s=(m_3 \pm m_4)^2$ \\ 
$t=0$ & $t=(m_1 \pm m_3)^2,$ & $t=(m_2 \pm m_4)^2$ \\ 
$u=0$ & $u=(m_1 \pm m_4)^2,$ & $u=(m_2 \pm m_3)^2$ \\ 
\end{tabular}
\caption{Kibble curve's tangents.}
\label{tab:kib_tangents}
\end{table}

\noindent
Let us now state some properties that will be useful in the more complicated case of the $2 \to 3$ scattering.\\
Constraints on Mandelstam variables coming from $\Delta_2\leq0$ are planes in the $(s,t,u)$ space; the constraint from the Principal Gram Determinant is more complicated.\\
The growing $u$ direction is from the bottom left to the upper right corner. Hence, when all bounds from \cref{tab:2_to_2_d2_constr} are respected, the central blob is not included in the $2 \to 2$ scattering phase space; instead, it is the $1 \to 3$ decay domain. The other 3 regions correspond to the $s, \ t \text{ or } u$ channel, respectively counter-clockwise from the bottom right corner. This means that every channel only picks one branch at a time of all the branches that are in \cref{tab:2_to_2_d2_constr}.\\

The $s$ region can be rotated into the $t$ one by plugging into the Kibble curve

\begin{equation}
s \leftrightarrow t \land m_2 \leftrightarrow m_3; \label{eq:s_t_rotation}
\end{equation}

\noindent
$s$ is rotated into $u$ via

\begin{equation}
s \leftrightarrow u \land m_2 \leftrightarrow m_4; \label{eq:s_u_rotation}
\end{equation} 

\noindent
$t$ is rotated into $u$ via

\begin{equation}
t \leftrightarrow u \land m_3 \leftrightarrow m_4. \label{eq:t_u_rotation}
\end{equation} 

\noindent
Therefore we can restrict to one channel and work out all other possibilities through symmetry.

\subsection{$1\to 3$ decay}
\begin{figure}[H]
\centering
\includegraphics[width=0.5\linewidth]{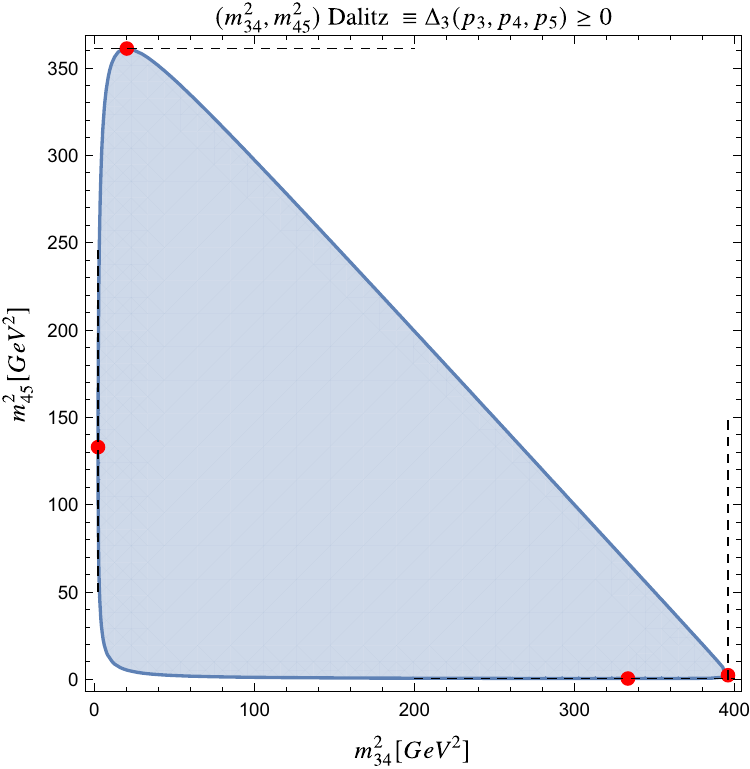}
\caption{$(m_{34}^2,m_{45}^2)$ Dalitz plot. Onto red dots the tangents (grey dashed lines) are either horizontal or vertical.}
\label{fig:d3dalitzequiv}
\end{figure}

The next step in complexity is provided by the $1 \to 3$ decay. A useful tool for its characterization are Dalitz plots \cite{dalitz}, representing the allowed phase space to two invariant masses. For example, if we consider three four-momenta $p_{3,4,5}$, a Dalitz plot is depicted in terms of $m_{34}^2:=(p_3+p_4)^2$ and $m_{45}^2:=(p_4+p_5)^2$. Generally they are used for resonance hunting: experimentalist collect events as data points that they draw in the $(m_{34}^2,m_{45}^2)$ plane, as in \cref{fig:d3dalitzequiv}. A vertical or horizontal band more populated than others can be the indication of a resonance. The populated area in a Dalitz plots corresponds to the area $\Delta_3(p_3,p_4,p_5)\geq 0$.

\subsection{$2\to3$ scattering}\label{sec:2_to_3_gds_kinematics}

In this section we express the $2\to3$ scattering or $1 \to 4$ decay phase space ${\cal P}_5$ in terms of GDs. Specifically, $\Delta_4$ is the Principal Gram Determinant of this problem and is the only ingredient of the phase space. Indeed the phase space volume is $\theta(-\Delta_4)$ and the phase space density is $\frac{1}{\sqrt{-\Delta_4}}$. We provide the derivation in \cref{sec:xsec}. Expressing ${\cal P}_5$ with GDs has many advantages: it is simpler and more compact, all the cross section distributions correspond to tabulated integrals and it exhibits an integrable divergence exactly where the $2\to3$ scattering is degenerate, i.e. $\Delta_4=0$. This easily points to the most frequent kinematic configurations. This formalism has also proven efficient in MC generators to speed up simulation \cite{madgraph_gd}.\\
In this section we define the Generalised Mandelstam variables (\cref{sec:gmv_def}) and deepen some considerations (\cref{d2_hyperbox}) already given in the previous section, which will be the foundation for performing the integrations of next section.

\subsubsection{Generalized Mandelstam variables}\label{sec:gmv_def}

In \cref{sec:gd_histo} we saw that 5 particles subject to four-momentum conservation can be described by 5 degrees of freedom. They can be the Mandelstam $s$ plus four other invariant masses $(s_1, \ s_2, \ t_1, \ t_2)$ that we call \textbf{Generalized Mandelstam variables (gMvs)}. They simply correspond to four-momenta scalar products $p_{ij}=p_i \cdot p_j$. In fact from now on we adopt the convention

\begin{convention}[The $2 \to 3$ scattering]
\begin{align}
&p_1+p_2 \to p_3+p_4+p_5,\\
&p_i^2=m_i^2,\\
&\vec{m}:=(m_1,m_2,m_3,m_4,m_5),
\end{align}
\end{convention}

\noindent
from which follow the definition

\begin{definition}[Generalized Mandelstam Variables]\label{def:gmvs}
\begin{equation}
\begin{split}
& s:= (p_1+p_2)^2,\\
& s_1:=(p_3+p_4)^2,\\
& s_2:= (p_4+p_5)^2,
\end{split}
\qquad \qquad
\begin{split}
& t_1:= (p_1-p_3)^2,\\
& t_2:=(p_2-p_5)^2.\\
\end{split}
\end{equation}
\end{definition}
In a $2\to2$ scattering with constant $s$ the only degree of freedom is $t$ but we can also define $u$ for our convenience. Analogously, in the $2\to3$ scattering $(s_1, \ s_2, \ t_1, \ t_2)$ are the only degrees of freedom but $u-$like variables can be defined

\begin{definition}[Auxiliary variables]\label{def:aux_vars}
\begin{equation}
\begin{split}
& u_1:=(p_3+p_5)^2,\\
& u_2:=(p_1-p_5)^2,\\
& u_3:=(p_2-p_3)^2,\\
\end{split}
\qquad \qquad
\begin{split}
& u_4:=(p_2-p_4)^2,\\
& u_5:=(p_1-p_4)^2.\\
&
\end{split}
\end{equation}
\end{definition}

\noindent
See \cref{shortcuts:sec:definitions} for conversion among scalar products, gMvs and auxiliary variables.\\ 

Using \cref{def:gmvs} and approximating the SM+ALP theory to a fully scalar theory, the matrix elements corresponding to \cref{fig:alp_chans} become

\begin{equation}
{\cal M}_s \to \dfrac{1}{s u_1}, \qquad {\cal M}_t \to \dfrac{1}{t_1 t_2}. \label{eq:ssm_m_s_t}
\end{equation}
respectively for the $s-$ and $t-$channel-like topologies.

\subsubsection{A system of second order inequalities}\label{d2_hyperbox}

If in the previous example we dealt with $\Delta_3$, let us upgrade now to $\Delta_4$, the Principal Gram Determinant of a $2 \to 3$ scattering:

\begin{equation}
\Delta_4=
\left|
\begin{smallmatrix}
m_1^2 & \frac{1}{2} \left(-m_1^2-m_2^2+s\right) & \frac{1}{2}
\left(m_1^2+m_3^2-t_1\right) & \frac{1}{2}
\left(-m_2^2+s-s_1+t_2\right) \\
\frac{1}{2} \left(-m_1^2-m_2^2+s\right) & m_2^2 & \frac{1}{2}
\left(-m_1^2+s-s_2+t_1\right) & \frac{1}{2}
\left(m_2^2+m_5^2-t_2\right) \\
\frac{1}{2} \left(m_1^2+m_3^2-t_1\right) & \frac{1}{2}
\left(-m_1^2+s-s_2+t_1\right) & m_3^2 & \frac{1}{2}
\left(m_4^2+s-s_1-s_2\right) \\
\frac{1}{2} \left(-m_2^2+s-s_1+t_2\right) & \frac{1}{2}
\left(m_2^2+m_5^2-t_2\right) & \frac{1}{2}
\left(m_4^2+s-s_1-s_2\right) & m_5^2
\end{smallmatrix}
\right| \label{eq:m4}.
\end{equation}

\noindent
We proceed to show some Gram Determinants properties that are responsible for the phase space shape. \Cref{eq:m4} shows that

\begin{theorem}[GDs as parabolas]\label{theo:gmvs_parabolas}
Gram Determinants of order $n$ are multivariate polynomials of the generalised Mandelstam variables of the same order.  In each of its monomial every variable has either order 0,1 or 2 and must recover the mass dimension of $2n$. Hence GDs can always be viewed as parabolas in one gMv whose coefficients may depend on other gMvs.
\end{theorem}

\begin{corollary}[Dimensions constraints]\label{corol:constr_dim}
The mass dimension of a $\Delta_n$ is

\begin{equation}
[\Delta_n]=n [s]= 2n.
\end{equation}
Call $c_2^{(x)}$ the $x^2$ coefficient in a GD of order $n$, for $x \in \{s,s_1,s_2,t_1,t_2\}$. Then

\begin{equation}
[c_2^{(x)} x^2]=[\Delta_n]=2 n=[c_2^{(x)}]+[x^2]=[c_2^{(x)}]+4,
\end{equation}

\noindent
from which

\begin{equation}
[c_2^{(x)}]=2n-4
\end{equation}

\noindent
It follows that the solution to $(-1)^{n+1}\Delta_n\geq0$ has dimension $2n-4$ in the space $(s,s_1,s_2,t_1,t_2)$. For example, the $\Delta_2$s give hyperplanar constraints (the tangents to the Kibble curve in \cref{sec:2_to_2_invitation}).
\end{corollary}

\noindent
In general conditions on lower dimensional GDs are contained in the conditions for higher dimensional $\Delta_n$. Therefore imposing $(-1)^{n+1}\Delta_n\geq0$ implies to satisfy all the $(-1)^{m+1}\Delta_m\geq0, \text{ for } m<n$. Specifically, when we said that ${\cal P}_5$ corresponds to $\Delta_4\leq0$ we implied that with $\Delta_4\leq0$ also come all the constraints on smaller degree GDs.\\
Consequently,

\begin{corollary}[Constraints on higher order $\Delta_n$ carve out a smaller phase space]\label{corol:phase_space_approx}
Consider a set of $n+1$ four-momenta $\{p_i\}_{i=1}^{n+1}$ with $\sum\limits_{i=1}^{n+1}p_1=0$. Its Principal Gram Determinant is $\Delta_n$, ${\cal R}_n$ is the region defined by

\begin{equation}
(-1)^{n+1} \Delta_n(\{p_i\}_{i=1}^n)\geq0,
\end{equation}

\noindent
and ${\cal R}_{n-1}$ is the region defined as the intersection of all possible $\Delta_{n-1}$ we can build on the four-momenta set:

\begin{equation}
\bigcap\limits_{k \in \{1, \dots n\}}[(-1)^n \Delta_{n-1}(\{p_i\}_{i=1}^n\setminus \{p_k\})\geq0].
\end{equation}
From \cref{corol:phase_space_approx}, ${\cal R}_{n-1}$ defines a bigger volume than ${\cal R}_n$. The latter gives a better approximation of the phase space than the former. 
\end{corollary}
This is why a process with fewer bodies in the final state is expected to have a bigger cross section than one with more particles:

\begin{corollary}[Fewer bodies means bigger phase space]
Consider a $n_i \to n_f$ process and a second process equal to the previous one but with an extra particle $X$ in the final state of mass $m_X\geq0$, so that all the shared particles have the same mass. Regardless of $m_X$, from \cref{corol:phase_space_approx} the second process has a smaller or equal phase space than the first one.
\end{corollary}

\begin{figure}[H]
\centering
\includegraphics[width=\linewidth]{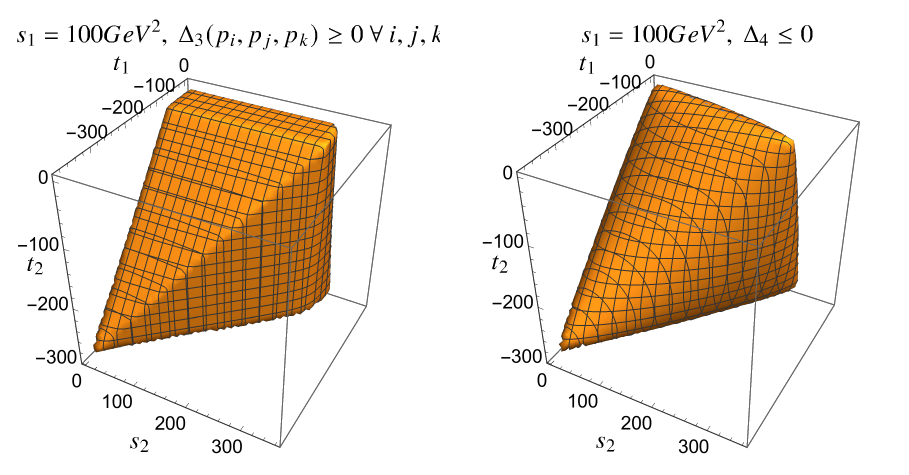}
\caption{Graphic example for \cref{corol:phase_space_approx}. Left: volume defined by $\Delta_3(p_i,p_j,p_k) \geq 0, \ i,j,k \in \{1,2,3,4,5\}$ in the $(s_1,t_1,t_2)$ phase space with $s=400  \GeV^2, \ s_1=100  \GeV^2, \ m_1=3  \GeV , \ m_2=2  \GeV , \ m_3=1  \GeV , \ m_4=0.5  \GeV , \ m_5=0.1  \GeV$. Right: same as left panel but here the volume is defined by $\Delta_4\leq0$ in a $2 \to 3$ scattering having .}
\label{fig:d3d4comapre}
\end{figure}

\begin{figure}[H]
\centering
\includegraphics[width=0.6\linewidth]{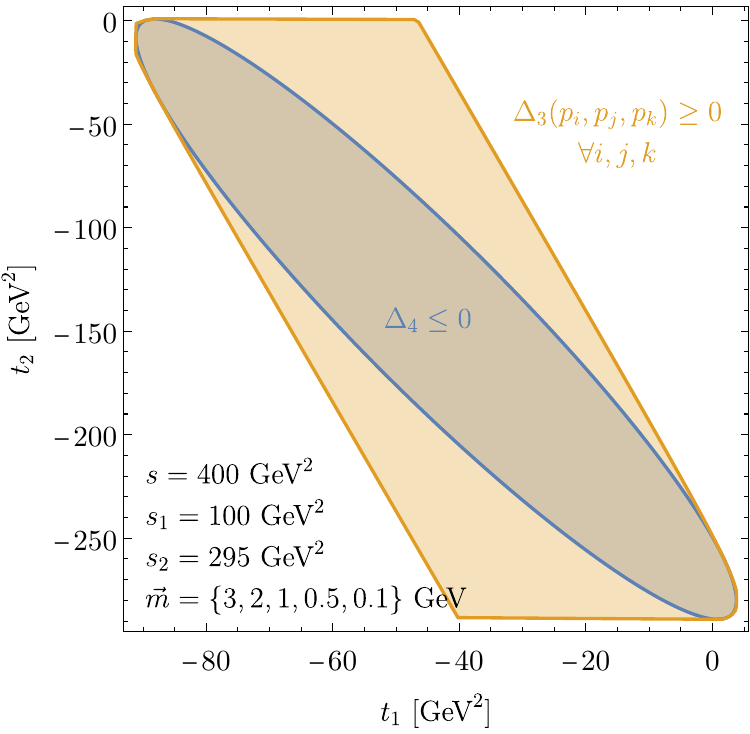}
\caption{$s_2=295 \GeV^2$ slice of \cref{fig:d3d4comapre}. We can see how ${\cal R}_4$ is entirely contained within ${\cal R}_3$, being at most tangent to it in some points.}
\label{fig:d3d4250slice}
\end{figure}

An example of \cref{corol:phase_space_approx} is depicted in \cref{fig:d3d4comapre} for a $2\to3$ scattering. The left panel depicts ${\cal R}_3$ appearing bigger than ${\cal R}_4$, on the right panel. This can be seen even more clearly in \cref{fig:d3d4250slice}, representing the slice $s_2=295 \GeV^2$ from \cref{fig:d3d4comapre}. ${\cal R}_4$ is entirely contained within ${\cal R}_3$, being at most tangent to it in some points.

Let us prove \cref{corol:phase_space_approx}. We will follow these steps: we consider a generic $\Delta_n, n>2$; as Kajantie and Lindblom state in \cite{2_to_3_finnici}, we remark that, via \cref{lemma:sum_one_mom}, we can choose one scalar product $p_l\cdot p_m, l \neq m$ and make it appear only in two symmetric positions of $M_n$. But then 

\begin{equation}
\Delta_n=a_2 (p_l\cdot p_m)^2+ a_1 (p_l\cdot p_m)+ a_0=0\label{eq:a_0_1_2}
\end{equation}

\noindent
is solved onto some special roots requiring for two $\Delta_{n-1}$ to have the proper sign. The demonstration being valid for any $n$ allows us to repeat it equally onto these new $\Delta_{n-1}$ down to the $\Delta_1$.\\
For this proof we will need some terminology:

\begin{convention}[Matrix elements and minors naming]
Let
\begin{itemize}
\item $A$ be a generic $n \times n$ symmetric matrix.
\item $a_i^j$ be the element at the $i^\text{th}$ row and the $j^\text{th}$ column.
\item $A_{\slashed{i_1}, \dots \slashed{i_k}}^{\slashed{j_1}, \dots \slashed{j_z}}$ be the minor obtained by erasing the  rows $(i_1, \dots i_k)$ and the columns $(j_1, \dots j_z)$ from $A$.
\item In our particular case $A$ is the Gram Matrix $M_n$ and we want to express it as a parabola as if $p_l\cdot p_m, l \neq m$ was the independent variable and all other scalar products were parameters.
\item In our proof there will be many sums. All of them are carried on indices among 1 and $n$. If as a sum limit we indicate, for example, $r \neq l$, we mean that we sum on $[1,n]\backslash\{l\}$.
\end{itemize}
\end{convention}

\noindent
For this proof we will use the Laplace expansion. Along the $i^\text{th}$ row

\begin{equation}
\det A = \sum\limits_{j=1}^n (-1)^{i+j} a_i^j A_{\slashed{i}}^{\slashed{j}}.
\label{laplace_determinant}
\end{equation}

\noindent
The Laplace expansion can be used twice. Pick two distinct rows $i,k$ and define

\begin{equation}
\det A = \sum\limits_{j=1}^n \sum\limits_{r \neq j} (-1)^{i+j+\phi(i,k)+\phi(j,r)} a_i^j a_k^r A_{\slashed{i},\slashed{k}}^{\slashed{j},\slashed{r}}.
\label{double_laplace_determinant}
\end{equation}

\noindent
where $\phi$ is needed to account for columns and rows cancellations. Here I provide one of its many possible definitions:

\begin{equation}
\phi(a,b)=
\begin{cases}
&b \text{ if } b<a,\\
&b-1 \text{ if } a<b,\\
&\text{error value if } a=b.
\end{cases}
\end{equation}

\noindent
Through \cref{double_laplace_determinant} it is easy to express $\Delta_n$ as a parabola in $x=p_l\cdot p_m$:

\begin{corollary}[$a_{lm}$ collection]
\begin{equation}
\begin{aligned}
\det A &=- a_l^m a_m^l A_{\slashed{l},\slashed{m}}^{\slashed{m},\slashed{l}}\\
& + \left[a_l^m \sum\limits_{k\neq l} (-1)^{l+k+ \phi(l,m)+\phi(m,k)} a_m^k A_{\slashed{l},\slashed{m}}^{\slashed{m},\slashed{k}}+
a_m^l \sum\limits_{j\neq m} (-1)^{m+j+ \phi(l,m)+\phi(j,l)} a_l^j A_{\slashed{l},\slashed{m}}^{\slashed{j},\slashed{l}}\right]\\
&+ \sum\limits_{j\neq m} \sum\limits_{k\neq j, l} (-1)^{l+j+m+k+ \phi(l,m)+\phi(j,k)} a_l^j a_m^k A_{\slashed{l},\slashed{m}}^{\slashed{j},\slashed{k}}.
\end{aligned}\label{eq:double_laplace}
\end{equation}
\end{corollary}

\noindent
Now in the first term in \cref{eq:double_laplace} we erased both the $l$th row and column and both the $m$th row and column. Applying \cref{eq:double_laplace} to our computation means that $A$ is a Gram Matrix of four-momenta $(p_1, \dots p_4)$ and $A_{\slashed{l},\slashed{m}}^{\slashed{m},\slashed{l}}$ is a symmetric Gram Determinant in which we erased the $l$th and the $m$th four-momentum.\\
We now provide some notation in order to identify each $a_{0,1,2}$ from \cref{eq:a_0_1_2} with the terms from \cref{eq:double_laplace}.

\begin{definition}[GD with erased entries]
Start from a set of independent four-momenta $\{p_i\}_{i=1}^n$ and its associated $\Delta_n$. $\Delta_{n-k}^{(\slashed{i_1}, \dots \slashed{i_k})}$ is $\Delta_n$ in which we eliminated the four-momenta with indices\footnote{This is not to be confused with the four-momentum in the position $i_1, \dots i_k$: a definition based on positions would have been meaningless given the GD symmetry under entries permutation.} $i_1, \dots i_k$. For example in the $2 \to 3$ scattering we will use $\Delta_4(p_1,p_2,p_3,p_5)$. $\Delta^{(\slashed{1})}$ is $\Delta_3(p_2,p_3,p_5)$, $\Delta^{(\slashed{2},\slashed{5})}=\Delta_2(p_1,p_3)$. In what follows we will be omitting the GD dimension when it will be obvious.
\end{definition}

\noindent
From the first row of \cref{eq:double_laplace}, the $a_{lm}^2$ coefficient in \cref{eq:double_laplace} is $-\Delta^{(\slashed{l}\slashed{m})}$.\\

The term in square brackets in \cref{eq:double_laplace} is made of two symmetric terms. For example in the first one we compute the $a_l^m$ cofactor as if $a_m^l$ was 0 and viceversa. Let us name this cofactor $V_{lm0}$.\\

Finally the last term in \cref{eq:double_laplace} is like computing the full GD with $a_l^m=a_m^l=0$.\\
In conclusion 

\begin{theorem}[Kajantie and Lindblom trick]\label{theo:k_l_trick}
Consider an $n \times n$ Gram Matrix $M$ and let
\begin{equation}
\Delta= a (M_{lm})^2 + b M_{lm} +c=a(M_{lm}-x_-)(M_{lm}-x_+).\label{eq:kl_problem_def}
\end{equation}

\noindent
with $M_{lm}=M_{ml}=p_l\cdot p_m$. You can show that

\begin{align}
& a= -\Delta^{(\slashed{l}\slashed{m})}= -\Delta^{(ml)} \label{eq:a_coef},\\
& b=\dfrac{V_{lm0}}{2}=\dfrac{V_{ml0}}{2}, \label{eq:b_coef}\\
& c= \Delta_n\big|_{x=0}, \label{eq:c_coef}\\
\end{align}
and that the $x_\pm$ from \cref{eq:kl_problem_def} are

\begin{equation}
x_\pm=\dfrac{V_{lm0}\pm\sqrt{\Delta^{(\slashed{l})}\Delta^{(\slashed{m})}}}{\Delta^{(\slashed{l}\slashed{m})}}\label{minor_trick_sol}.
\end{equation}
\end{theorem}

\noindent
Kajantie and Lindblom state the latter for a generic $n$ without demonstrating it nor pointing to any reference for it. A proof by induction is possible but too lengthy to report here.\\

In conclusion, for $x_\pm$ to be real we need $\Delta^{(\slashed{l})}\Delta^{(\slashed{m})}\geq0$, i.e. $\Delta^{(\slashed{l})}$ and $\Delta^{(\slashed{m})}$ must have the same sign. The sign can be recovered from \cref{signs_bounds_theorem}. This concludes our proof.


\subsubsection{Roughest Phase space approximation: the circumscribed hypercube}\label{subsec:branch_selection}

In this section we will focus on bounds coming from all possible $\Delta_2$s. Practically we will inscribe the physical region for a $2\to3$ process into a hyperbox in the $(s_1,s_2,t_1,t_2)$ space\footnote{See \cref{def:gmvs} for the Generalized Madelstam variables definition.} so that it will be clear that the process is bounded from all sides. Consequently all possible divergences of cross sections may only come from the matrix element diverging somewhere on the physical region.\\

Consider

\begin{equation}
\Delta_2(p_1,p_2)\leq 0 \to s \geq(m_1+m_2)^2 \cup s\leq (m_1-m_2)^2.
\end{equation}

\noindent
Since all $\Delta_2(p_i,p_j)$ need to be $\leq0$, instead of performing computations every time, we may also permute the indices of the four-momenta (and those of the masses with them). In this way we can generalize to 

\begin{equation}
(p_i \pm p_j)^2\geq (m_i+m_j)^2 \cup (p_i \pm p_j)^2\leq (m_i-m_j)^2.
\label{delta_2_inequalities}
\end{equation}

\noindent
Of the two branches we must pick

\begin{equation}
\begin{cases}
(p_i+p_j)^2\geq (m_i+m_j)^2,\\
(p_i-p_j)^2\leq (m_i-m_j)^2,
\end{cases}
\end{equation}

\noindent
if we want

\begin{equation}
p_i \cdot p_j = E_i E_j - \cos \theta_{ij} |\vec{p}_i||\vec{p}_j|\geq 0,\label{eq:base_scattering_selection}
\end{equation}
which is the case since $E_i \geq |\vec{p}_i| \geq 0$ and $\cos \theta_{ij}\leq 1$. Notice how choosing \cref{eq:base_scattering_selection} is the same thing of choosing the $s$ branch in the Kibble curve in \cref{fig:Kibble}.\\

We finally provide two more constraints to complete the gMvs hyperbox: $\Delta_2(p_1+p_2,p_3) \leq 0$ is solved by $\{s_2 \leq (\sqrt{s}-m_3)^2\} \cup \{s_2 \geq (\sqrt{s}+m_3)^2\}$. Since we know that $s_2\leq s$ we can only accept the solution $s_2 \leq (\sqrt{s}-m_3)^2$. The same reasoning works for $\Delta_2(p_1+p_2,p_5) \leq 0 \to s_1 \leq (\sqrt{s}-m_5)^2$. These last two bounds are generally valid for any scattering. Consider in fact $p_{i,j}$ two final state four-momenta and name $P=(P^0,\vec{P})$ the sum of all initial state four-momenta. Go to the frame in which all the remaining final four-momenta are at rest

\begin{equation}
\sum\limits_{k \neq i,j} p_k=(M, \vec{0}).
\end{equation}

\noindent
Then

\begin{equation}
(p_i+p_j)^2=\left(P-\sum\limits_{k \neq i,j} p_k\right)^2=(\sqrt{s}-M)^2-2M(P^0-\sqrt{s})\leq (\sqrt{s}-M)^2.
\end{equation}

\noindent
The last inequality is valid since $P^0\geq \sqrt{s}$, as $P^2=(P^0)^2-\vec{P}^2=s\geq0$. Let us summarize this result in

\begin{theorem}[Further bound on $s$-like variables]

In a $n \to m$ process, let the initial state invariant mass be $\sqrt{s}$, take $i,j$ in the final state and let the invariant mass of all other final particles be $M$. Then

\begin{equation}
(p_i+p_j)^2\leq (\sqrt{s}-M)^2.
\end{equation}
\end{theorem}

Something similar can be said for $t$-type variables. We provide for simplicity a demonstration for a $2 \to 3$ process. Consider

\begin{equation}
u_2=(p_1-p_5)^2=(p_2-p_3-p_4)^2=m_2^2+s_1-2 p_2 \cdot (p_3+p_4).
\end{equation}

\noindent
$p_2 \cdot (p_3+p_4)$ is biggest if $p_5=(0,\vec{0})$, but then $p_3+p_4=p_1+p_2$. From this we can derive the following theorem

\begin{theorem}[Further bounds on $t-$like variables]
Consider a $2 \to 3$ scattering and define generalized Mandelstam variables as in \cref{def:gmvs}. The following bounds hold:
\begin{align}
\begin{split}
&t_1>m_1^2+s_2-s,\\
&t_2>m_2^2+s_1-s,\\
&u_2>m_1^2+s_1-s,
\end{split}
\qquad \qquad
\begin{split}
&u_3>m_2^2+s_2-s,\\
&u_4>m_2^2+u_1-s,\\
&u_5>m_1^2+u_1-s.
\end{split}
\end{align}
\end{theorem}

To wrap up we collect here all the useful bounds for the gMvs when the positive branch for $s$ is chosen:

\begin{theorem}[gMvs are bounded from above and from below]\label{theo:gmv_box}
\begin{subequations}
\begin{align}
&(m_3+m_4)^2 \leq s_1 \leq (\sqrt{s}-m_5)^2,\\
&(m_4+m_5)^2 \leq s_2 \leq (\sqrt{s}-m_3)^2,\\
&m_1^2+s_2-s<t_1 \leq (m_1-m_3)^2,\\
&m_2^2+s_1-s<t_2 \leq (m_2-m_5)^2.
\end{align}
\label{d2_bounds}
\end{subequations}
\end{theorem}

\section{$2 \to 3$ cross section and distributions}\label{sec:xsec}
We are now ready to solve \cref{eq:dd_jacobian}. Indeed, using \cref{s_f_ij_ik}

\begin{equation}
a^2 \left[1-\left(\dfrac{b}{a}\right)^2\right]= \dfrac{-4 \Delta_4}{ \lambda(s,m_1^2,m_2^2)},
\end{equation}

\noindent
($\lambda$ is defined in \cref{def:tri_lambda}). We can now transform from polar coordinates to gMvs so to exploit the simpler expression of $\Delta_4$ we just found in \cref{d2_hyperbox}. Therefore we will trade the polar coordinate integration measure from \cref{def:pol_measure} for a Lorentz invariant one:

\begin{align}
& d \mu_P = d s_1 d s_2 d t_1 d t_2 |J| =: d \mu_L |J|,\\
& |J|=\dfrac{1}{16 \lambda(s,m_1^2,m_2^2)},
\end{align}

\noindent
where we defined $d \mu_L$ as the volume element for the gMvs. In conclusion the $2 \to 3$ scattering cross section is expressed in terms of Lorentz invariant quantities as

\begin{align}
&\lin:=\lambda(s,m_1^2,m_2^2),\label{def:lin}\\
& \sigma_0=\dfrac{1}{32(2 \pi)^4 \sqrt{\lin}} \frac{1}{\text{flux}},\\
& {\cal D}:= \dfrac{\theta(-\Delta_4)}{\sqrt{-\Delta_4}}, \label{2_to_3_density}\\
&\sigma = \sigma_0 \int d \mu_L |{\cal M}|^2 {\cal D}, \label{gmv_cross_section_full_integral}
\end{align}
where ${\cal D}$ is often referred to as the scattering density.\\

We can now integrate, for example, in $s_2$, using Kajantie and Lindblom's trick. Specifically we will use \cref{theo:k_l_trick} with $s_2$ as independent variable supposing the matrix element is not depending on it. Notice that this is true in our fully scalar theory \cref{eq:ssm_m_s_t}. If this was not the case it would still be possible to perform the integration as the generic integral

\begin{equation}
I_x^{(n)}:=\int_{x_-}^{x_+} dx \dfrac{x^n}{\sqrt{(x-x_-)(x_+-x)}}
\end{equation}
is solved to

\begin{equation}
I_x^{(n)}=\pi x_-^n  \,_2F_1\left(\frac{1}{2},-n,1,1,1-\frac{x_+}{x_-}\right)
\end{equation}
with assumptions $n \in \mathbb{Z}, \ x_\pm \in \mathbb{R}$. They come from asking that $I_x^{(n)}$ is an integration on some gMv for a physical matrix element. $_2F_1$ is the hypergeometric function.\\

Sticking to the $s_2$ independent case, let us define the $s_2$ integration as

\begin{equation}
I_{s_2}:=\int d s_2 \dfrac{\theta(-\Delta_4)}{\sqrt{-\Delta_4}}.\label{def:is2}
\end{equation}

\noindent
We can express $\Delta_4$ as a polynomial in $s_2$ and call its roots $s_2^\pm$ (we will always be using this convention from now on):

\begin{equation}
\Delta_4= a s_2^2+ b s_2 + c=a(s_2-s_2^-)(s_2-s_2^+). \label{eq:d4_s2_parabola}
\end{equation}

\noindent
By using \cref{minor_trick_sol} we can deduce that

\begin{equation}
\begin{aligned}
b^2 - 4 a c &=64 \Delta_3^{(\slashed{3})}\Delta_3^{(\slashed{5})}=64 \Delta_3(p_1,p_2-p_5,p_5)\Delta_3(p_1,p_2-p_5,p_3)\\
&= 4 {\cal G}_1(s, t_2, s_1, m_2^2, m_1^2 m_5^2) {\cal G}_2(t_1, s_1, t_2, m_3^2, m_1^2, m_4^2).
\end{aligned}
\end{equation}

\noindent
For this we deduce that further integration will take place in the region identified by ${\cal G}_{1,2}\leq0$. \Cref{gmv_cross_section_full_integral} becomes

\begin{equation}
\sigma=\sigma_0 \int d s_1 d t_1 d t_2 \theta(-{\cal G}_1)\theta(-{\cal G}_2) I_{s_2}.\label{eq:intermediate_is2}
\end{equation}

\noindent
${\cal G}_{1,2}$ will be crucial in the following sections hence we explicitly write them

\begin{align}
& {\cal G}_1=\dfrac{M_e^2 s_1^2}{2}-s_1 \left(\dfrac{s \ t_2}{2}+M_e^4\right)+ \dfrac{1}{2}(M_e^6-3 M_e^2 s \ t_2 + s \ t_2(s+t_2))\label{eq:g1},\\
& \begin{aligned}
{\cal G}_2&=\dfrac{t_1 s_1^2}{2}-\dfrac{1}{2}s_1 t_1(M_a^2+2 M_e^2+t_2-t_1)\\
&+ \dfrac{1}{2}(M_a^4 M_e^2+M_e^4 t_1+ M_e^2 t_2(t_2-t_1) + M_a^2(t_1 t_2-M_e^2(t_1+2 t_2)))\label{eq:g2}.
\end{aligned}
\end{align}

\noindent
Lastly, we can solve \cref{def:is2} using \cref{eq:d4_s2_parabola}:

\begin{equation}
\begin{aligned}
I_{s_2}&:= \int\limits_{s_2^-}^{s_2^+} \dfrac{d s_2 }{\sqrt{-\Delta_4}}=\dfrac{\pi}{\sqrt{a}}.
\end{aligned} \label{is2}\\
\end{equation}

From \cref{theo:k_l_trick}:

\begin{equation}
a=\dfrac{\lambda(s_1,t_2,m_1^2)}{16}.
\end{equation}

\noindent
A concrete example of matrix element not depending on $s_2$ is that of the $t$ channel, for which ${\cal M}_t=\frac{1}{t_1^2 t_2^2}$. Using \cref{eq:intermediate_is2,is2}, the total cross section reads

\begin{equation}
\sigma_t=\sigma_0 \pi \int \dfrac{d t_1 d t_2 d s_1}{t_1^2 t_2^2} \dfrac{\theta(-{\cal G}_1)\theta(-{\cal G}_2)}{\sqrt{\lambda(s_1,t_2,m_1^2)}}.\label{eq:d_sigma_t_dt1dt2ds1}
\end{equation}

\subsection{$\ddv{\sigma_t}{t_2}{s_1}$}\label{section:dst_dt2_ds1}

Let us continue the computation we started in \cref{eq:d_sigma_t_dt1dt2ds1} and integrate $t_1$ away, so to obtain a fully analytical expression for $\ddv{\sigma_t}{t_2}{s_1}$ of the $t-$channel. The first step is finding the integration domain, the second is integrating.\\ The integration domain is easily found because only ${\cal G}_2$ depends on $t_1$:

\begin{equation}
{\cal G}_2=\frac{s_1^2}{2}(t_1-t_1^-)(t_1-t_1^+) \leq 0,\label{eq:g2_ineq}
\end{equation}

\noindent
realised onto

\begin{equation}
t_1^- \leq t_1 \leq t_1^+<0.
\end{equation}

\noindent
In \cref{eq:g2_ineq}, \cref{eq:g2} is a polynomial in $t_1$ with roots $t_1^\pm$:

\begin{equation}
\begin{aligned}
&2 s_1 t_1^\pm= s_1 t_2-s_1^2+M_a^2(s_1-t_2)+M_e^2(M_a^2 + 2 s_1+t_2)-M_e^4+\\
&\pm\sqrt{\lambda(s_1,M_e^2,M_a^2)\lambda(s_1,t_2,M_e^2)}
\end{aligned}\label{eq:tpm_sols}.
\end{equation}

\noindent
Notice how \cref{eq:tpm_sols} respects the form predicted in \cref{theo:k_l_trick}. Moreover $t_1^+=0$ if $t_2=M_a^2$, which is impossible since $t_2\leq0$ always. Therefore the dangerous $t_1=0$ subspace is always excluded. The subspace $t_2=0$ is excluded too for symmetry.\\
At this point the double cross section distribution domain must satisfy both 

\begin{equation}
{\cal G}_1=\dfrac{M_e^2}{2}(s_1-s_1^-)(s_1-s_1^+)\leq0
\end{equation}

\noindent
and the constraints coming from asking that the solutions in \cref{eq:tpm_sols} are real:

\begin{equation}
\begin{cases}
&s_1^+ \leq (M_e+M_a)^2 \leq s_1 \leq s_1^- \leq (\sqrt{s}-M_e)^2,\\
&\min t_2 \leq t_2 \leq \max t_2,
\end{cases}
\end{equation}

\noindent
with

\begin{align}
& \min t_2=(M_e+M_a)^2+M_e^2-\dfrac{s+(M_e+M_a)^2-M_e^2}{2}+\dfrac{\sqrt{\lin \lambda(s,(M_a+M_e)^2,M_e^2)}}{2s}, \label{eq:min_t2_chew_low}\\
& \max t_2=(M_e+M_a)^2+M_e^2-\dfrac{s+(M_e+M_a)^2-M_e^2}{2}-\dfrac{\sqrt{\lin \lambda(s,(M_a+M_e)^2,M_e^2)}}{2s}. \label{eq:max_t2_chew_low}
\end{align}

\noindent
The results are in agreement with ref. \cite{2_to_3_finnici} (we remind that in this case $\lin=\lambda(s,M_e^2,M_e^2)$). The boundary solutions in Lorentz invariant variables \cref{eq:min_t2_chew_low,eq:max_t2_chew_low} correspond to boundaries in polar coordinates:

\begin{align}
&\max E_{3,5}=\dfrac{s-2 M_a M_e-M_a^2}{2 \sqrt{s}},\\
&\max t_1(E_3,c_3)=t_1(\max E_3,1),\\
&\max t_2(E_5,c_5)=t_1(\max E_5,-1)= \max t_1.\\
\end{align}

\begin{figure}[!h]
\includegraphics[width=\linewidth]{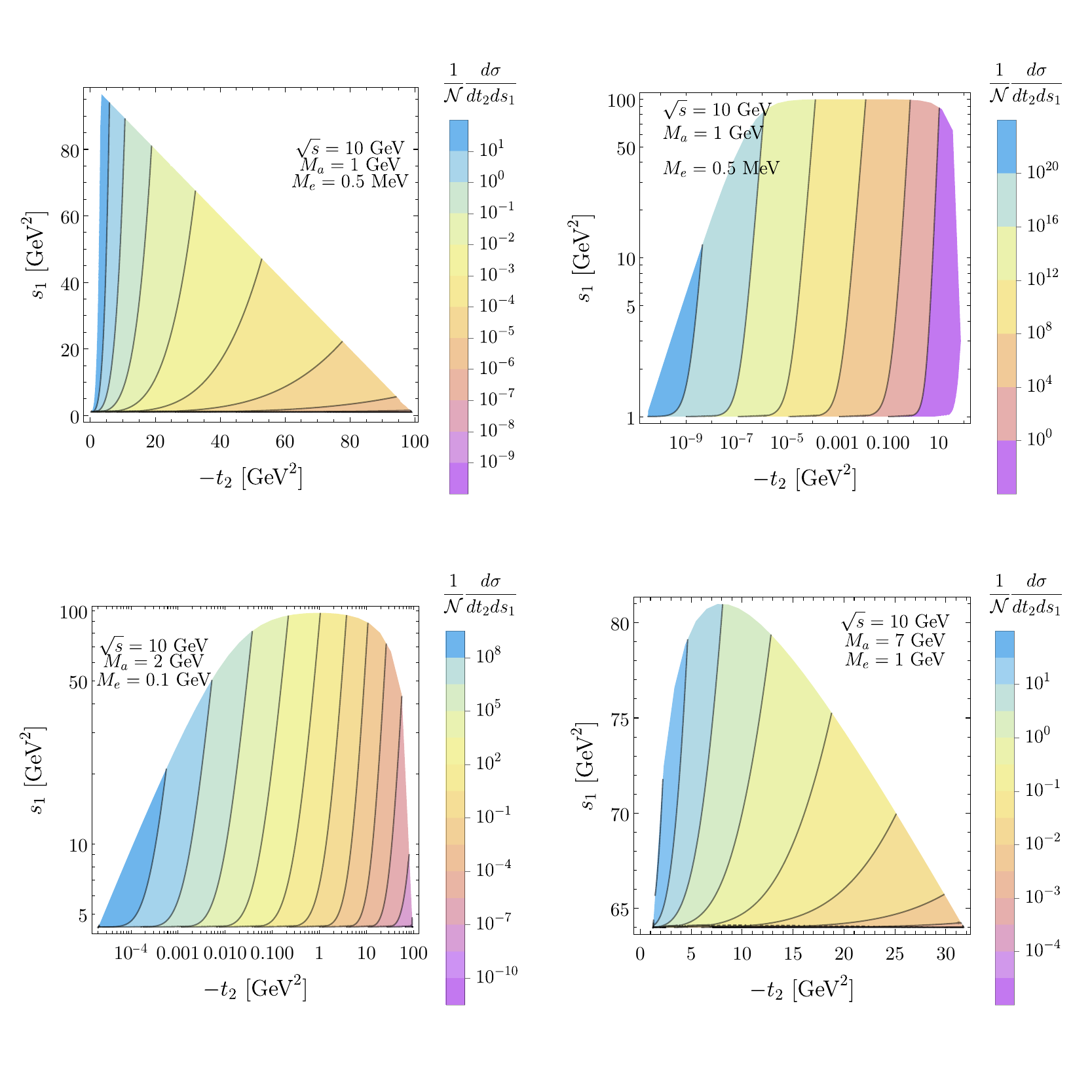}
\caption{Normalized $\ddv{\sigma_t}{t_2}{s_1}$ for different values of the $M_{a,e}$. In every panel: $\sqrt{s}=10$ GeV, ${\cal N}$ is the cross section distribution evaluated on $s_1=70 \text{ GeV}^2, t_2=-10 \text{ GeV}^2$. Top row: $M_a=1$ GeV, $M_e=0.5$ MeV. Bottom row, left: $M_a=2$ GeV, $M_e=0.1$ GeV; bottom row, right: $M_a=7$ GeV, $M_e=1$ GeV.}
\label{fig:2d_t_distributions}
\end{figure}

This concludes our domain considerations. We can now proceed and integrate in $t_1$: 

\begin{equation}
\begin{aligned}
\sigma_t&=\sigma_0 \pi \int \dfrac{d t_2 d s_1}{t_2^2} \dfrac{\theta(-{\cal G}_1)}{\sqrt{\lambda(s_1,t_2,m_1^2)}}\int \dfrac{d t_1}{t_1^2} \theta(-{\cal G}_2)\\
&=:\dfrac{\sigma_0 \pi}{M_e^2} \int d t_2 \ d s_1 \ \theta(-{\cal G}_1)  \ \rho(s_1,t_2,M_a,M_e),\
\end{aligned}
\end{equation}

\noindent
with

\begin{equation}
\rho(s_1,t_2):=\dfrac{\sqrt{\lambda(s_1,M_a^2,M_e^2)}}{t_2^2 \ (t_2-M_a^2)^2}.
\end{equation}
The cross section distribution is depicted in \cref{fig:2d_t_distributions} for different masses choices. In all cases the cross section peaks at small $|t_2|$, but the effect is more intense when the masses are sufficiently small that $\max t_2 \sim 0$. In this case, only a small fraction of $\sqrt{s}$ must be converted to masses, so that final state energies can be similar to initial state energies and make small virtualities. Now the kinematic configurations where the cross section distribution is maximized are those that will be realized more often at colliders so it is crucial to analytically predict maxima. While we know that the maximum is at small $|t_2|$ because the matrix element peaks there, the maximum can not be exactly on the boundary as the boundary cuts the cross section to zero. These two competing effects produce the maximum. A more precise derivation of the maximum is given in the next section.

\subsubsection{Maximization of $\ddv{\sigma_t}{t_2}{s_1}$}
It is useful to study this double differential cross section, particularly the phase space areas making it maximal so to deduce the most relevant kinematic configurations. When evaluated on the physical domain, $\ddv{\sigma_t}{t_2}{s_1}$ is an always positive function so it is indifferent to study $\rho$ or $\rho^2$. The gradient is always strictly positive

\begin{align}
& \dv{\rho^2}{s_1}=2\dfrac{s_1-M_a^2-M_e^2}{(M_a^2-t_2)^4 (-t_2)^4},\\
& \dv{\rho^2}{t_2}=\rho^2\dfrac{4(M_a^2-2 t_2)}{(M_a^2-t_2) (-t_2)}.
\end{align}

\begin{figure}[!h]
\centering
\includegraphics[width=0.75\linewidth]{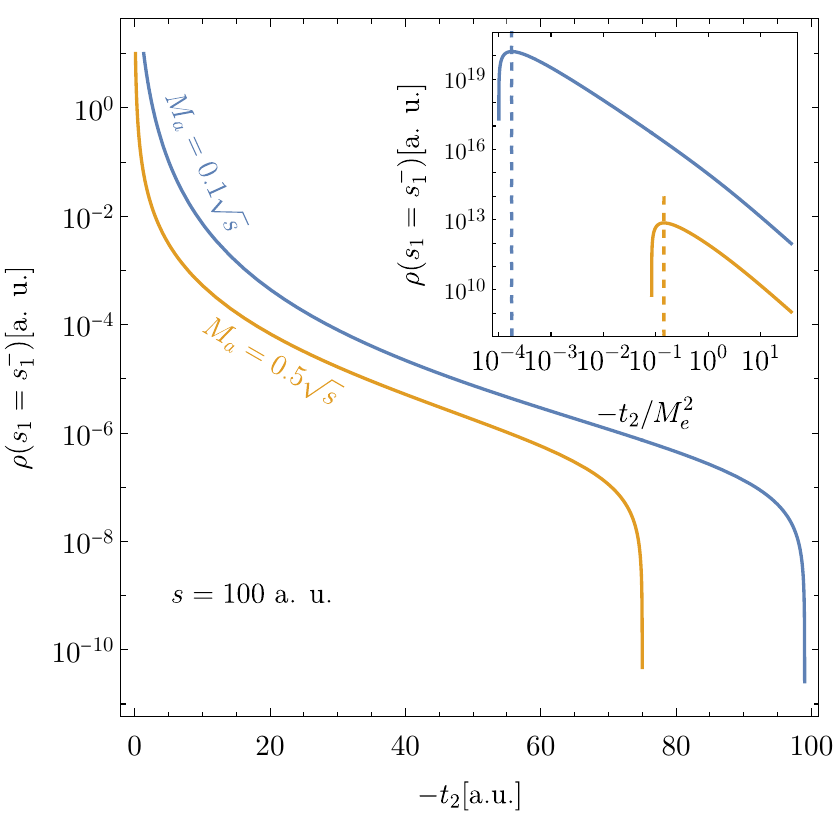}
\caption{$\rho(s_1=s_1^-,t_2)$ for $M_a=0.1 \sqrt{s}$ and $M_a=0.5\sqrt{s}$ in blue and orange respectively. The inset shows a zoomed version of the same function to exhibit the maximum next to the phase space closure. Dashed lines correspond to $t_2=\tilde{t_2}$ following \cref{eq:tilde_t2}. $M_e=5 \cdot 10^{-5} \sqrt{s}$.}
\label{fig:dsdt2ds1_on_s1max}
\end{figure}

\noindent
The gradient annihilates on the point $(s_1=(M_a+M_e)^2,t_2=M_a^2/2)$, therefore in the limit $M_a, M_e \to 0$ the maximum is attained on the phase space boundary, otherwise the gradient annihilates on an unphysical point. Both $M_a$ and $M_e$ are preventing the divergence. For finite $M_{a,e}$ the absolute maximum of $\rho$ is to be searched for onto the boundary of the function domain. In \cref{fig:dsdt2ds1_on_s1max} we represent $\rho(s_1=s_1^-)$ as a function of $t_2$ for different $M_a$ values.

\begin{equation}
\frac{\dd \rho}{\dd t_2} \bigg|_{s_1=s_1^-}=\left(\frac{\partial \rho}{\partial t_2} +\frac{\partial \rho}{\partial s_1} \frac{\dd s_1^-}{\dd t_2}\right)_{s_1=s_1^-}.\label{eq:rho_der_on_bound}
\end{equation}
This derivative can annihilate within the physical phase space, therefore $\rho$ has a maximum. 
The proof of the existence of the maximum depends on the signs in \cref{eq:rho_der_on_bound}. If $\frac{\dd s_1^-}{\dd t_2}\geq 0$ then the $\rho$ derivative is positive. But $\frac{\dd s_1^-}{\dd t_2}$ can be negative if $t_2>-M_e(\sqrt{s}-2M_e)$. An example point where $\frac{\dd s_1^-}{\dd t_2}<0$ is $(s_1,t_2)=(\min s_1, \max t_2)=: \pi_-$. This is a physical point because $\min s_1=\min s_1^-=(M_a+M_e)^2$ is attained on $\max t_2>-M_e(\sqrt{s}-2M_e)$. We can show that the total derivative is negative at $\pi_-$. Indeed, $\dv{\rho^2}{t_2}=0$ on $s_1=(M_a+M_e)^2$, so

\begin{equation}
\frac{\dd \rho}{\dd t_2} \bigg|_{s_1=(M_a+M_e)^2}=\left(\frac{\partial \rho}{\partial s_1} \frac{\dd s_1^-}{\dd t_2}\right)_{s_1=(M_a+M_e)^2}<0.
\end{equation}
Thus, the maximum exists for a value of $t_2$ close to its maximum. 

We could maximize $\rho(s_1^-,t_2)$ but it is a complicated task so we first perform an expansion in the small quantities of this problems. Say $\rho(s_1^-,t_2)$ is maximal at some $t_2=\bar{t}_2 \simeq \min t_2 \simeq 0$.  We expand in small $\dfrac{t_2}{s}$ and small $\dfrac{M_e}{\sqrt{s}}$. The position of the expansion maximum is denoted by $\tilde{t}_2$ and gives the following satisfactory approximation of $\bar{t}_2$:

\begin{equation}
\tilde{t}_2=\dfrac{M_e^2}{4 s(s- M_a^2 )}\left[9s^2-8 M_a^2 s -4 M_a^4-(3 s - 2 M_a^2)\sqrt{9 s^2-4 M_a^2 s +4 M_a^4}\right] \simeq \bar{t}_2.\label{eq:tilde_t2}
\end{equation}
Notice how $\tilde{t}_2 \to0$ both for $M_a\to0$ and $M_e\to 0$. For a comparison between $\bar{t}_2$ and $\tilde{t}_2$ see the inset of \cref{fig:dsdt2ds1_on_s1max}.

We conclude by verifying the kinematic configurations corresponding to $s_1=\min s_1$ and $t_2=\max t_2$. For $s_1$ to be minimal we need the ALP and the electron to have the same velocity. In fact, go to the rest frame of $p_3$ and let $\vec{\beta}_4$ and $\gamma_4$ be the ALP velocity and Lorentz factor respectively. Then

\begin{equation}
s_1=(p_3+p_4)^2=m_3^2+m_4^2+2 m_3 m_4 \gamma_4 \geq (m_3+m_4)^2.
\end{equation}

\noindent
The minimum is attained when ALP and electron have the same velocity. In particular if in the CoM they are both at rest, then $E_5^*$ is maximal.\\

Maximal $t_2$ can be attained onto maximal $E_5$ and $\cos \theta_5\to-1$. In fact, in the rest frame of $p_2$, with $\vec{\beta}_5$ and $\gamma_5$ the final positron velocity and Lorentz factor respectively,

\begin{equation}
t_2=(p_2-p_5)^2=m_2^2+m_5^2-2 m_2 m_5 \gamma_5 \leq (m_2-m_5)^2.
\end{equation}

\noindent
The maximum is attained when initial and final positron have the same velocity. Since their masses are the same, the whole four-momenta will be equal.

\subsection{$\dv{\sigma_t}{t_2}$}\label{d_sigma_t_d_t2}
We may finally integrate $s_1$ away so to obtain a single differential cross section. It is remarkable that the GD formalism allows us to compute a distribution that in general is not easy to recover analytically and yet is notably useful at experiments. We obtain

\begin{equation}
\begin{aligned}
\dv{\sigma_t}{t_2}&= \dfrac{\sigma_0 \pi}{M_e^2} \dfrac{1}{t_2^2(t_2-M_a^2)^2}\bigg[\dfrac{1}{2}\sqrt{\lambda(s_1^-,M_a^2,M_e^2)}(s_1^--(M_a^2+M_e^2))\\
&+2(M_a M_e)^2 \log\left(\dfrac{s_1^--(M_a^2+M_e^2)-\sqrt{\lambda(s_1^-,M_a^2,M_e^2)}}{2 M_a M_e}\right)\bigg].
\end{aligned}
\label{dsigmat_dt2}
\end{equation}

\noindent
Since the $s_1$ integration was performed on the interval $[(M_a+M_e)^2,s_1^-]$, there is no $s_1^+$ in \cref{dsigmat_dt2}. We plot it in \cref{fig:dist_1d}, right panel.\\

To understand the magnitude of $\dv{\sigma_t}{t_2}$ we can expand it in small $M_e/\sqrt{s}$

\begin{equation}
\dv{\sigma_t}{t_2}=\sigma_0 \pi \frac{(-M_a+s+t_2)^2}{2 M_e^2t_2^2(M_a^2-t_2)^2}+{\cal O}\left(\frac{M_e}{\sqrt{s}}\right)
\label{eq:ds_dt2_expand}
\end{equation}
Notice how this expansion is only valid if $M_e/\sqrt{s}$ is the only small parameter. Indeed in \cref{fig:dist_1d} we show the expansion in dashed lines: while it captures with great accuracy the behavior at high $|t_2|$, it fails at small $|t_2|$. If we were to integrate up to $t_2=0$, \cref{eq:ds_dt2_expand} would diverge. This shows how virtuality cutoffs must be kept at every stage of the integration, especially if the cutoff to the final $e^\pm$ energies has to be $M_e$. Indeed, for a fully visible final state at a collider the energy cutoff is given by the energy threshold of the detector $E_0$. At a lepton collider this is typically way larger than the masses of the leptons in the beams. This energy hierarchy disentangles the matrix element scale $M_e/\sqrt{s}$ from the phase space scale $E_0/\sqrt{s}$, making it possible to expand the matrix element without affecting the phase space integration. The error induced on cross sections and distributions is small. The same can not be said for scatterings with invisible final state particles. Let us consider a VBF with final state electrons lost along the beam pipe, corresponding to the small $|t_{1,2}|$ region we are studying. This constraint is selecting exactly the phase space region that is close to divergence, i.e. the region where most of the events will be generated. There are no cuts that we could impose to avoid this problematic region, therefore both the matrix element and the phase space scales are $M_e/\sqrt{s}$. The smaller this scale is, the more likely it is that MC integration softwares are not able to reproduce the correct cross sections and cross section distributions. A solution can be using an Equivalent Vector Approximation (EVA) \cite{eva}. While the EVA provides a powerful simplification in VBF channels, it has some disadvantages: it can be complicated to understand for which processes and under which assumptions the EVA has an acceptable error with respect to the whole computation \cite{eva_fail}. Therefore the non approximated GD treatment serves at least as an independent validation of approximate methods, if not as the only reliable estimation of cross sections and cross section distributions.


\subsection{$\ddv{\sigma_s}{u_1}{s_2}$}\label{d_sigma_s_d_u1_d_s2}
In this section we provide an example of $s$ channel computation that will be useful for comparison with respect to the $t$ channel. We will follow the same path of the previous section. In this case the matrix element is $|{\cal M}_s|^2=\dfrac{1}{s^2 u_1^2}$, where $s_1$ and $u_1$ are related by \cref{eqs:from_mad_to_aux_and_back}, so that $d s_1= d u_1$. Then

\begin{equation}
\sigma_s=\sigma_0 \int \dfrac{d s_2 d u_1 d t_1 d t_2}{s^2 u_1^2} \dfrac{\theta(-\Delta_4)}{\sqrt{-\Delta_4}}.
\end{equation}

\noindent
We use again \cref{theo:k_l_trick} with independent variable $t_2$ in position, $(2,4), \ (4,2)$ of the gram matrix $M_4(p_1+p_2,p_2,p_3,p_5)$. Using also \cref{is2}

\begin{equation}
\int\limits_{t_2^-}^{t_2^+} \dfrac{d t_2}{\sqrt{-\Delta_4}}=\dfrac{4 \pi}{\sqrt{\lambda(s,s_2,m_3^2)}}.
\end{equation}

\noindent
Then

\begin{equation}
\sigma_s=4 \sigma_0 \pi \int \dfrac{d s_2 d u_1 d t_1}{s^2 u_1^2} \dfrac{\theta(-{\cal G}_1)\theta(-{\cal G}_2)}{\sqrt{\lambda(s,s_2,m_3^2)}},
\end{equation}

\noindent
with

\begin{align}
&{\cal G}_1:={\cal G}(s,t_1,s_2,m_1^2,m_2^2,m_3^2),\\
&{\cal G}_2:={\cal G}(u_1,s_2,m_5^2,s_,m_4^2,m_3^2).
\end{align}

\noindent
Proceeding with the $t_1$ integration, the inequality:

\begin{equation}
{\cal G}_1=\dfrac{s}{2}(t_1-t_1^-)(t_1-t_1^+) \leq 0
\end{equation}
is solved by

\begin{equation}
t_1^-\leq t_1 \leq t_1^+.
\end{equation}
Then

\begin{equation}
\begin{aligned}
\sigma_s&=4 \sigma_0 \pi \int \dfrac{d s_2 d u_1}{s^2 u_1^2} \dfrac{\theta(-{\cal G}_2)}{\sqrt{\lambda(s,s_2,m_3^2)}}\int\limits_{t_1^-}^{t_1^+} dt_1\\
&= 4 \sigma_0 \pi \int \dfrac{d s_2 d u_1}{s^2 u_1^2} \dfrac{\theta(-{\cal G}_2)}{\sqrt{\lambda(s,s_2,m_3^2)}}\dfrac{\sqrt{\lambda(s,m_1^2,m_2^2)}\sqrt{\lambda(s,s_2,m_3^2)}}{s}\theta(\lambda(s,s_2,m_3^2))\\
&=\sigma_0 \pi \sqrt{\lin} \int \dfrac{d s_2 d u_1 }{s^3 u_1^2} \theta(-{\cal G}_2)\theta(\lambda(s,s_2,m_3^2)),
\end{aligned}\label{eq:dss_ds2_du1}
\end{equation}

\noindent
where $\lin$ is defined in \cref{def:lin}.\\

In this case too we may keep going and integrate over $s_2$ so to recover a single derivative cross section distribution. What are the integration bounds? Due to the masses of our problem, it is always verified that if ${\cal G}_2 \leq 0$ then also $\lambda(s,s_2,m_3^2)\geq0$, therefore, with our usual notation,

\begin{equation}
{\cal G}_2 \leq 0 \text{ if } s_2^-\leq s_2 \leq s_2^+.
\end{equation}

\noindent
The double cross section distribution $\ddv{\sigma_s}{s_2}{u_1}$ from \cref{eq:dss_ds2_du1} does not depend on $s_2$ so we only have to integrate $\theta(-{\cal G}_2)$ away:

\begin{equation}
\int d s_2 \theta(-{\cal G}_2)= \theta(\Lambda(u_1))(s_2^+-s_2^-)=\theta(\Lambda(u_1))\dfrac{\sqrt{\Lambda(u_1)}}{u_1},
\end{equation}
where

\begin{equation}
\Lambda(u_1):=\lambda(u_1,m_3^2,m_5^2)\lambda(u_1,s,m_4^2).\label{eq:Lambda}
\end{equation}

\begin{figure}
\centering
\includegraphics[width=\linewidth]{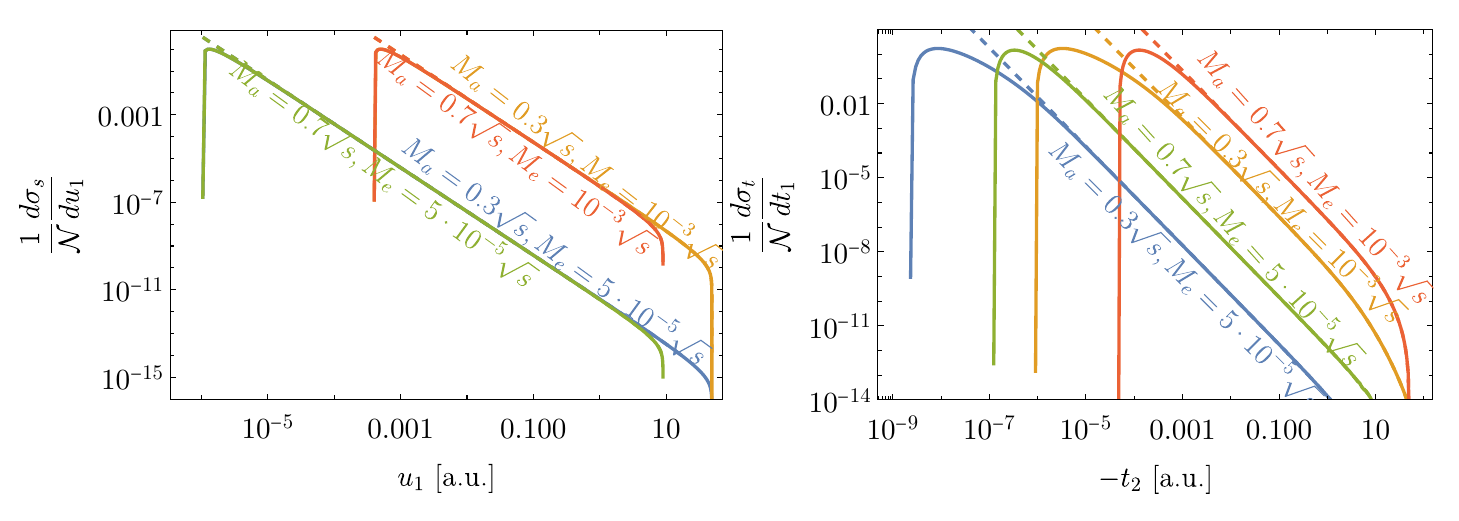}
\caption{One dimensional cross section distributions for different matrix elements and on different choices of masses. In both panels $\sqrt{s}=10$ GeV and dashed lines correspond to approximations in the largest order in $M_e/\sqrt{s}$. Left: cross section distribution for the $s$ channel matrix element with respect to $u_1$ as per \cref{eq:dss_du1}; curves are normalized by the distribution evaluated on $u_1=u_1^{\max}$, the maximizing point. Right: cross section distribution for the $t$ channel matrix element with respect to $t_2$ as per \cref{dsigmat_dt2}; curves are normalized by the distribution evaluated on $t_2=\tilde{t}_2$, an approximation of the maximizing point we derived in \cref{eq:tilde_t2}.}
\label{fig:dist_1d}
\end{figure}

\noindent
In conclusion

\begin{equation}
\dv{\sigma_s}{u_1}= \dfrac{4 \sigma_0 \pi}{s^3 u_1^3} \sqrt{ \lin \ \Lambda(u_1)} \theta(\Lambda(u_1))\label{eq:dss_du1}.
\end{equation}
This distribution is plotted in \cref{fig:dist_1d}, right panel. It is easy (but too lengthy to write here) to find the maximum of \cref{eq:dss_du1} because it is simply done by solving a third order polynomial equation: we can derive the distribution with respect to $u_1$ and find that $u_1^{\max}$ that annihilates the derivative. If $M_e\to0$, one of the roots of the derivative goes to 0, making the function always decreasing in the physical $u_1$ interval. 

The integrand is not diverging in the limit of small $M_e/\sqrt{s}$ if the lower bound to $u_1$ is kept finite:

\begin{equation}
\dv{\sigma_s}{u_1}= \dfrac{4 \sigma_0 \pi \sqrt{\lambda(u_1,s,M_a^2)}}{s^2 u_1^2} + {\cal O} \left(\frac{M_e^2}{s}\right)\label{eq:dss_du1_expand}.
\end{equation}
In the limit of massless and collinear final electrons, this cross section will diverge. The approximation for small $M_e/\sqrt{s}$ is depicted in \cref{fig:dist_1d}, left panel, as dashed lines and we can see the always decreasing behavior we described earlier.\\ 
Expansions make it particularly easy to see how much more important the $t-$ channel can be with resect to the $s-$channel. Indeed, for finite $t_2, u_1$ we can compare \cref{eq:dss_du1_expand} (with an $s^2$ in the denominator) with \cref{eq:ds_dt2_expand} (with an $M_e^2$ in the denominator). This result aligns with the fact that future, very high energy colliders will measure way more VBF events rather than production events \cite{mu_smash}, penalized by inverse powers of $s$.

\begin{figure}
\centering
\includegraphics[width=0.6\linewidth]{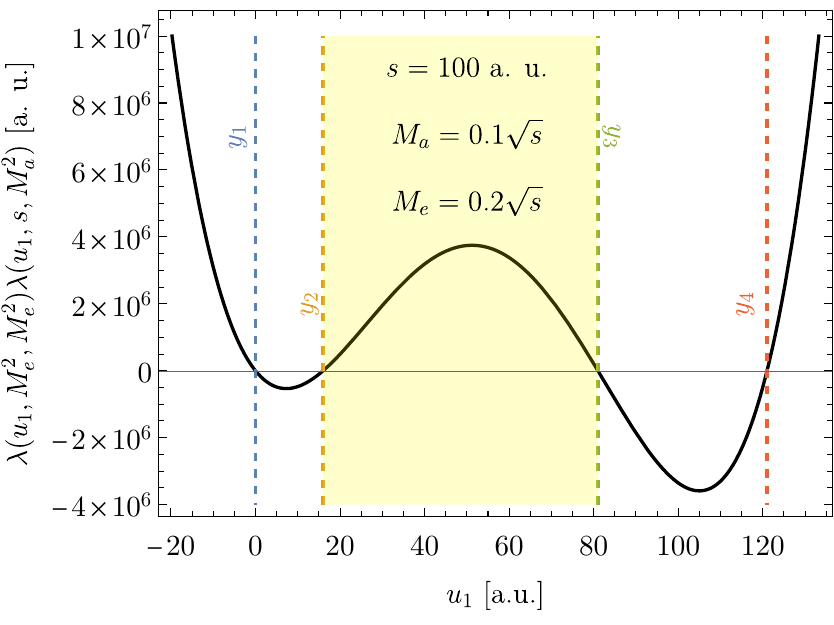}
\caption{$\Lambda$ function from \cref{eq:Lambda} on an example string of parameters $\sqrt{s}=10, M_a=1, M_e=2$ a.u. Vertical dashed lines indicate the $y_{1,2}$, the roots of $\lambda(u_1,m_3^2,m_5^2)$, in blue and orange respectively; $y_{3,4}$ are the roots of $\lambda(u_1,s,m_4^2)$ and are indicated by the green and red dashed lines respectively. The yellow area highlights the physical $u_1$ interval $y_2 \leq u_1 \leq y_3$.}
\label{fig:u14roots}
\end{figure}

It is possible to analytically integrate \cref{eq:dss_du1}. $\Lambda(u_1)$ is the product of two parabolas in $u_1$. At least in our specific problem, their roots are ordered. Call $y_{1,2}$ the roots of $\lambda(u_1,m_3^2,m_5^2)$ and $y_{3,4}$ the roots of $\lambda(u_1,s,m_4^2)$ as polynomials in $u_1$ 

\begin{equation}
\lambda(u_1,m_3^2,m_5^2)=0 \text{ if } u_1=y_{1,2}; \qquad \lambda(u_1,s,m_4^2)=0 \text{ if } u_1=y_{3,4}.
\end{equation}
We plot them for example masses in \cref{fig:u14roots}.

The following chain of inequalities is true

\begin{equation}
0 \leq (m_3-m_5)^2=y_1 \leq y_2 \leq y_3 \leq s \leq y_4, \ y_4=(\sqrt{s}+m_4)^2 \neq0.
\end{equation}

\noindent
To integrate \cref{eq:dss_du1} we need to satisfy $\Lambda(u_1) \geq 0$, which is true onto $y_2 \leq u_1 \leq y_3$, therefore the integration extrema are $y_{2,3}$. We may exploit the fact that $y_4 \neq0$ and divide $u_1$ and all the $y_i$ by $y_4$, so to rewrite the integral in an easier way:

\begin{equation}
\sigma_s \propto y_4 \int\limits_{x_2}^{x_3} \dfrac{\sqrt{x-x_1} \sqrt{x-x_2} \sqrt{x-x_3} \sqrt{x-1}}{x^2} d x.
\end{equation}

\noindent
with $\frac{u_1}{y_4}=:x$ and $x_i:=\frac{y_i}{y_4}$. These kinds of integrals are tabulated and available in symbolic computation softwares like Mathematica \cite{mathematica_12}. A case that is of interest for us is $y_1= 0$, for which

\begin{equation}
\begin{aligned}
&\sigma_s(x_1=0)=\dfrac{4 \sigma_0 \pi \sqrt{\lambda_i}}{s^3}\bigg[\frac{2}{3 x_2}\frac{r_{42}}{r_{32} \sqrt{x_3}} \Sigma ~ E(\kappa)+\\
&+\dfrac{r_{43} r_{42}}{r_{32} x_3}(x_2+x_3) F\left(\alpha,\frac{1}{\kappa}\right)- \dfrac{r_{42} }{x_3^{3/2}}\Sigma ~	K(\kappa)-3 \frac{x_2}{r_{23}}\Pi\left(x_3,\alpha,\frac{1}{\kappa}\right)\bigg]
\end{aligned},\label{eq:full_sigma_s}
\end{equation}

\noindent
where

\begin{align}
& r_{ij}:=\sqrt{x_i-x_j}, \ i>j,\\
& \kappa:=\dfrac{x_3-x_2}{x_3(1-x_2)},\\
& \Sigma:= x_2+x_3+x_2 x_3\\
&\alpha:=\arcsin(\sqrt{\kappa}).
\end{align}

\noindent
$E$ and $\Pi$ are incomplete elliptic integrals respectively of the second and third kind while $K$ is the complete integral of the first kind.

\subsection{$\ddv{\sigma_t}{t_1}{t_2}$}\label{sec:dst_dt1_dt2}

$\ddv{\sigma_t}{t_1}{t_2}$ is the most difficult distribution as in this case the kinematic functions are entangled in a complicated way. However not only we can solve it, $\ddv{\sigma_t}{t_1}{t_2}$ also is the distribution from which we can learn the most about the $t$ channel. In fact in this section we will first characterize the $(t_1,t_2)$ phase space (\cref{subsec:t1_t2_ps}); we will demonstrate that it gets tremendously close to the matrix element poles but keeps them outside of its boundaries. The two dimensional distribution is found in \cref{eq:ds_dt1_dt2}.

\subsubsection{The $s_1$ integration}\label{subsec:t1_t2_ps}
We can start this computation from \cref{eq:d_sigma_t_dt1dt2ds1} (i.e. from $\dfrac{d^3 \sigma_t}{d s_1 d t_1 d t_2}$) and integrate $s_1$ away:

\begin{equation}
\int d s_1 \theta(-{\cal G}_1) \theta(-{\cal G}_2)= \int\limits_{f(t_1,t_2)}^{g(t_1,t_2)} d s_1.
\end{equation}

\noindent
We shall find the extrema $f$ and $g$.\\
The two Heaviside thetas prescribe that a point in the space $(s_1,t_1,t_2)$ is inside the physical region if both ${\cal G}_{1,2}\leq 0$. Separately, ${\cal G}_{1,2}\leq 0$ define the volumes ${\cal V}_{1,2}$. Together the two inequalities define the intersection of these volumes. Since $s_1, t_1$ and $t_2$ all have upper and lower bounds, there is a parallelepiped in which the solving volume is inscribed, therefore the intersection of ${\cal V}_{1,2}$ defines the convex volume ${\cal V}$. 

\begin{figure}[!h]
\centering
\includegraphics[width=0.475\linewidth]{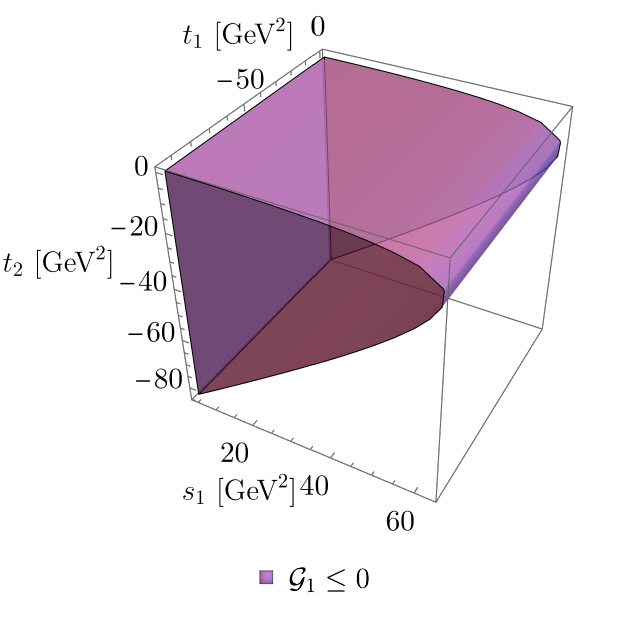}
\includegraphics[width=0.475\linewidth]{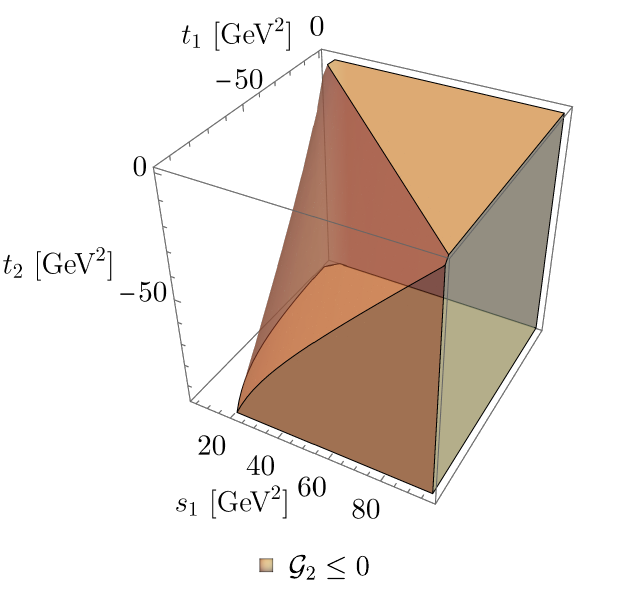}
\includegraphics[width=0.475\linewidth]{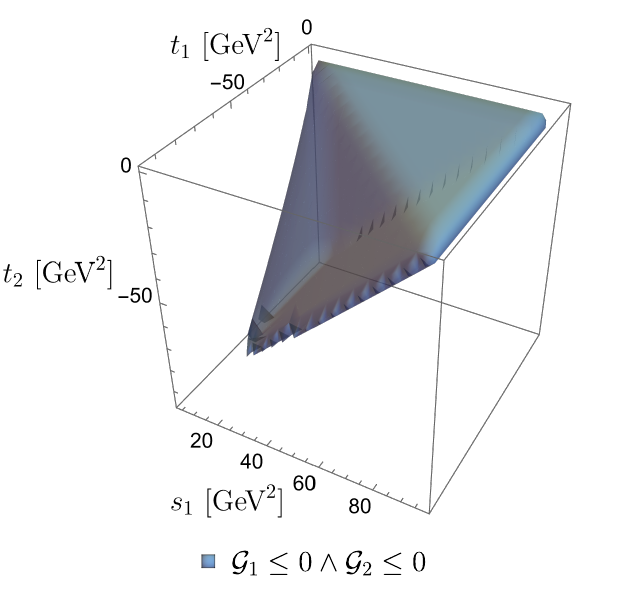}
\caption{Phase space defined by ${\cal G}_1\leq 0$ (\cref{eq:g1}) or ${\cal G}_2\leq 0$ (\cref{eq:g2}) for $\sqrt{s}=10, M_a=1, M_e =5.11 \cdot 10^{-4}$ GeV. Top left: ${\cal G}_1\leq 0$. Top right: ${\cal G}_2\leq 0$. Bottom: ${\cal G}_1\leq 0$ and ${\cal G}_2\leq 0$.}
\label{fig:g1_g2_volumes}
\end{figure}

Specifically, ${\cal G}_1$ (\cref{eq:g1}) is a convex parabola in $s_1$ (the $s_1^2$ coefficient is $\frac{M_e^2}{2}$). This means that ${\cal V}_1$ is convex. The equality ${\cal G}_1=0$ can only be achieved on the boundary $\partial {\cal V}_1$, but not every point of $\partial {\cal V}_1$ belongs to $\partial {\cal V}$. This happens when the zeros of ${\cal G}_1$ fall outside the physical region. Still, $\partial {\cal V}_1$ is a 1D continuous curve, see \cref{fig:g1_g2_volumes}, top left panel.

\noindent
${\cal G}_2$ (\cref{eq:g2}) is a convex parabola in $t_1$ (the $t_1^2$ coefficient is $\dfrac{s_1}{2}$). Then there is a convex set of points $ {\cal V}_2$ in the $(s_1,t_2,t_1)$ space where ${\cal G}_2\leq0$. The border of this volume $ \partial {\cal V}_2$ is defined by the physical roots of ${\cal G}_2=0$ and the boundaries on $s_1, t_1$ and $t_2$. $ \partial {\cal V}_2$ is a convex 2D surface in the 3D $(s_1,t_2,t_1)$ space, see \cref{fig:g1_g2_volumes}, top right panel.\\
Since we want both ${\cal G}_1\leq0$ and ${\cal G}_2\leq0$ at the same time, we will only retain the intersection ${\cal V}:= {\cal V}_1 \cap  {\cal V}_2$. Notice how if ${\cal V}= \emptyset$ our process is unphysical. 
We evince that the boundary is to be found onto 

\begin{equation}
\begin{cases}
{\cal G}_1=0,\\
{\cal G}_2=0.
\end{cases}
\label{gg_sys}
\end{equation}
i.e. when ${\cal G}_1$ and ${\cal G}_2$ share roots. Since ${\cal V}_1$ and ${\cal V}_2$ are convex, ${\cal V}$ is convex. ${\cal V}$ is a volume in the 3D $(s_1,t_2,t_1)$ space. Depending on the shape of ${\cal V}_1$ and ${\cal V}_2$, the border of ${\cal V}$ is the union of subsets of the borders of the parent sets. $\partial {\cal V}$ is a 2D curve in the 3D $(s_1,t_2,t_1)$ space. To find the $s_1$ integration extrema and the $(t_1,t_2)$ phase space boundary, consider \cref{fig:g1_g2_volumes}, bottom panel, and imagine to scan it from left to right with a plane $s_1=s_0$. On this plane you will see a convex shape. The union of all these shapes is the $(t_1,t_2)$ phase space ${\cal V}_t$. Its boundary $\partial{\cal V}_t$ is a 1D curve.\\
Let us work out $\partial{\cal V}_t$. Kajantie and Lindblom \cite{2_to_3_finnici} do it for a generic case but can not provide a generic curve for $\partial{\cal V}_t$ since it heavily depends on the masses of the $2 \to 3$ scattering. They only manage to recover some salient loci of the curve and then proceed to specialise to simple configurations. We extend the treatment of \cite{2_to_3_finnici} as we do not put any of the scattering masses to 0, but take all of them to be $M_e$ except $m_4$, that is the ALP mass. This allows us to explicitly derive $\partial{\cal V}_t$ as a function of $t_2$.\\ 
Call $s_{1i}^\pm$ the roots of ${\cal G}_i$ when solved for $s_1$. Explicitly

\begin{align}
&s_{11}^\pm=\dfrac{2M_e^4+s ~ t_2\pm \sqrt{\lambda(s,M_e^2,M_e^2)\lambda(t_2,M_e^2,M_e^2)}}{2M_e^2},\label{eq:s11_pm}\\
&s_{12}^\pm=\dfrac{M_a^2 t_1 +2 M_e^2 t_1 - t_1^2+ t_1 t_2\pm \sqrt{\lambda(s,M_e^2,M_e^2)\lambda(t_1,t_2,M_a^2)}}{2 t_1}.\label{eq:s12_pm}
\end{align}

\noindent
We need to check whether $s_{1i}^\pm$ fall inside the physical domain of definition of $s_1$. On the boundary at least one of the following is satisfied:

\begin{align}
& s_{11}^+=s_{12}^- \label{1p2m_boundary},\\
& s_{11}^-=s_{12}^+ \label{1m2p_boundary},\\
&t_1=\max t_1= (m_1-m_3)^2 \label{t1_boundary},\\
&t_2=\max t_2= (m_2-m_5)^2 \label{t2_boundary},\\
&\lambda(t_1,t_2,m_4^2)=0 \label{lambda_boundary}.
\end{align}

\noindent
\Cref{lambda_boundary} would need at least one between $t_1$ and $t_2$ to be positive, which is unphysical when $m_1=m_3, m_2=m_5$. If \cref{t1_boundary,t2_boundary} were realized, we would have VBF divergences onto the phase space boundary. They can be discarded easily for our scattering as, if we plug $t_{1,2}=0$ in \cref{gg_sys}, we get

\begin{align}
&{\cal G}_1(t_2=0)=\dfrac{M_e^2}{2}(s_1-M_e^2)^2,\\
&{\cal G}_2(t_1=0)=\dfrac{M_e^2}{2}(t_2-M_a^2)^2.
\end{align}

\noindent
The solutions of both are non physical and this is enough for excluding \cref{t1_boundary,t2_boundary}. Then for each $t_2$ either \cref{1p2m_boundary} or \cref{1m2p_boundary} must be true. Since our set of masses is particularly easy we can decide which. Let us begin this task by just inspecting ${\cal G}_{1,2}$ as parabolas in $s_1$:

\begin{align}
& {\cal G}_1=\dfrac{M_e^2 s_1^2}{2}-s_1 \left(\dfrac{s ~ t_2}{2}+M_e^4\right)+ \dfrac{1}{2}(M_e^6-3 M_e^2 s ~ t_2 + s ~ t_2(s+t_2)),\\
& \begin{aligned}
{\cal G}_2&=\dfrac{t_1 s_1^2}{2}-\dfrac{1}{2}s_1 t_1 (M_a^2+2 M_e^2+t_2-t_1)\\
&+ \dfrac{1}{2}(M_a^4 M_e^2+M_e^4 t_1+ M_e^2 t_2(t_2-t_1) + M_a^2(t_1 t_2-M_e^2(t_1+2 t_2))).
\end{aligned}
\end{align}

\noindent

The $s_1^2$ and $s_1$ coefficients in ${\cal G}_1$ are positive while the shift is negative. We can already tell that it will have only one real positive root. About ${\cal G}_2$, instead, we can only tell the $s_1^2$ coefficient is negative, hence there may be 0,1 or 2 acceptable roots. If there are 0, the process is unphysical. If there are 1 or 2 the process is physical but the least work would be done if only 1 root is acceptable. We can show that this is luckily our case since not even the verge of the parabola is allowed in the phase space. In fact if you solve $\dv{{\cal G}_2}{s_2}=0$ and plug the solution in ${\cal G}_2$, you can never satisfy ${\cal G}_2\leq0$. We conclude that there is only one physical root for ${\cal G}_1$ and one physical root for ${\cal G}_2$, but at this stage we don't know which. Nonetheless, we discovered that only one between \cref{1p2m_boundary,1m2p_boundary} can be true. We may decide which root is acceptable even by looking at them at some specific point being particularly easy for us. Also, expanding in small $\dfrac{M_e}{\sqrt{s}}=: \epsilon$ can not affect the reasoning above. The two expanded ${\cal G}_1=0$ solutions are

\begin{align}
&s_{11}^-= s+t_2+ {\cal O}(\epsilon^2),\\
&s_{11}^+=\epsilon^{-2} t_2-s-t_2+ {\cal O}(\epsilon^2).
\end{align}

\noindent
Since $s_{11}^+$ is always negative, we will keep $s_{11}^+$.\\
To double check, we can inspect $s_{12}^\pm$ too; we will adopt a different strategy and expand on the line $t_2=t_1(1+\epsilon)$ which we are sure belongs to the physical region: 

\begin{equation}
s_{12}^\pm = \dfrac{M_a^2}{2}+M_e^2\pm\dfrac{\sqrt{4 M_e^2-t_1}\sqrt{M_a^4-4M_a^2 t_1}}{2 \sqrt{-t_1}} + {\cal O}(\epsilon).
\end{equation}

\begin{figure}[!h]
\centering
\includegraphics[width=0.7\linewidth]{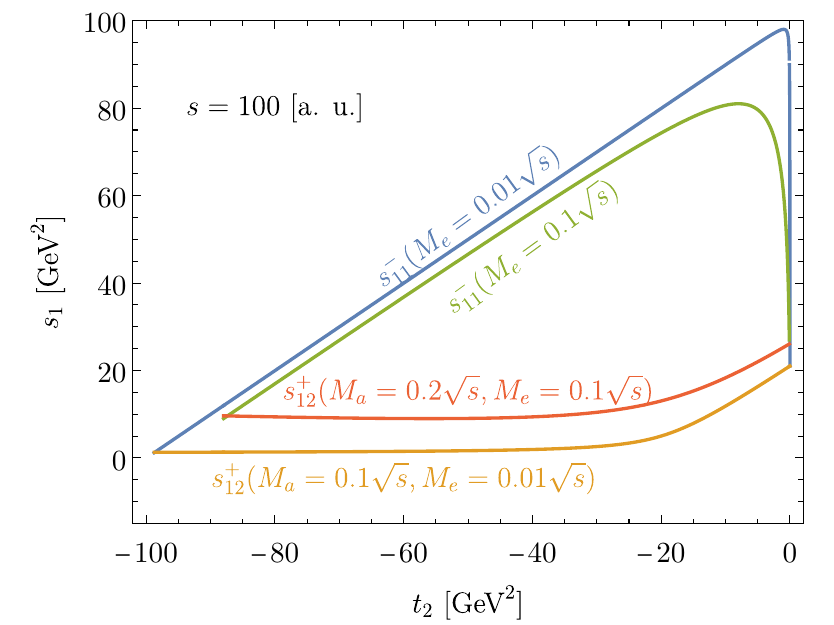}
\caption{$s_{11}^-$ is everywhere concave and $s_{12}^+$ is everywhere convex in $t_1$ and they intersect on the domain boundary, therefore, inside the physical region, $s_{11}^-\geq s_{12}^+$. Here we give a specific realization with $s= 100 \GeV^2, \ M_a= 1 \GeV, \ M_e=0.5 \GeV, \ t_1=-20 \GeV^2$.}
\label{fig:convexconcave}
\end{figure}

\noindent
The fact that we can only retain $s_{12}^+$ gives us the second confirmation that we must discard \cref{1p2m_boundary} and retain \cref{1m2p_boundary}. $t_2$ will live on a segment and, for each $t_2$, $t_1$ will be bounded form above and from below: $t_1^{\min}(t_2) \leq t_1 \leq t_1^{\max}(t_2)$. $t_1^{\min}$ and $t_1^{\max}$ are the two branches of the solution to $s_{11}^-=s_{12}^+$, a quadratic equation in $t_1$.\\ 
Finally, for each $t_1$ and $t_2$, in order to perform the $s_1$ integration we need to establish the biggest among $s_{11}^-$ and $s_{12}^+$. $s_{11}^-$ is everywhere concave and $s_{12}^+$ is everywhere convex in $t_2$ and they intersect on the domain boundary, see \cref{fig:convexconcave} for an example. Therefore, inside the physical region

\begin{equation}
s_{11}^-\geq s_{12}^+.
\end{equation}
The concavity of $s_{11}^-$ comes from the sign of its second derivative with respect to $t_{2}$:

\begin{equation}
\dvv{s_{11}^-}{t_2}=-\frac{2 M_e^2 \sqrt{\lin}}{ \lambda(t_2,M_e^2,M_e^2)^{3/2}}<0 \forall t_2 \in {\cal P}_5.
\end{equation}
The convexity of $s_{12}^-$ comes from the sign of its Hessian with respect to $t_{1,2}$:

\begin{equation}
H=\frac{\left(M_a^4 M_e^2+M_a^2 \left(2 M_e^2 (t_1-t_2)-t_1^2\right)+M_e^2 (t_1-t_2)^2\right)^2}{t_1^3 \left(4
M_e^2-t_1\right) \left(M_a^4-2 M_a^2 (t_1+t_2)+(t_1-t_2)^2\right)^2}<0 \forall t_{1,2} \in {\cal P}_5.
\end{equation}

In conclusion, for matrix elements $|{\cal M}(t_1, t_2)|^2$ not depending on $s_1$ and $s_2$

\begin{equation}
\dfrac{d \sigma}{d t_1 d t_2} \propto \dfrac{\sigma_0 |{\cal M}(t_1, t_2)|^2 \pi^2}{4 \sqrt{\lambda(s,m_1^2,m_2^2)}} \ln \left(\dfrac{s_{11}^--t_2-m_1^2-\sqrt{\lambda(s_{11}^-,t_2,m_1^2)}}{s_{12}^+-t_2-m_1^2-\sqrt{\lambda(s_{12}^+,t_2,m_1^2)}}\right), \label{eq:ds_dt1_dt2}
\end{equation}

\noindent
where $s_{11}^-$ and $s_{12}^+$ are defined in \cref{eq:s11_pm,eq:s12_pm}. The two dimensional cross section distribution with constant matrix element is plotted in \cref{fig:phi_dt1_dt2} for different choices of $M_e$. We defined $\Phi_3$ the cross section with constant matrix element. The scalar matrix element cross section distribution is plotted in \cref{fig:dst_dt1_dt2} for different choices of $M_e$. Let us remark that the exact knowledge of the phase space boundary allows us to compute the correct cross section: for example in the bottom right panel if we had considered $\min |t_1|=\min |t_2|=10^{-4}$ we would have lost orders of magnitude in cross section.\\
Notice that $\dd \sigma_t/ \dd t_2$ was already obtained in \cref{section:dst_dt2_ds1}. It is possible to integrate \cref{eq:ds_dt1_dt2} at least numerically in $t_1$ to check that we can obtain the same result but this gives us the opportunity to remark that some successions of integrations, although arriving at the same result, can be way more cumbersome than others. If the aim is obtaining $\dd \sigma_t/ \dd t_2$ or $\dd \sigma_t/ \dd t_1$ we suggest the procedure from \cref{section:dst_dt2_ds1}.\\
\Cref{eq:ds_dt1_dt2} may seem somewhat obscure; the $0^\text{th}$ order expansion in $M_e$ of the argument of the logarithm can be clarifying:

\begin{equation}
\dfrac{- 2 t_1 t_2}{s(t_1+t_2-M_a^2+\sqrt{\lambda(t_1,t_2,M_a^2)}}.
\end{equation}
It is now easier to notice two properties of this cross section distribution: i) it is symmetric under $t_1 \leftrightarrow t_2$; ii) the total cross section grows with $s$. Indeed, the phase space volume grows with $s$ and the cross section distribution is positive definite on the phase space; moreover, $\min |t_{1,2}|$ are smaller for larger $s$. The cross section is larger the smaller $|t_{1,2}|$ can become due to the dependence $\sim\log(t_1 t_2)/t_1^2t_2^2$. 

\begin{figure}[!h]
\centering
\includegraphics[width=0.75\linewidth]{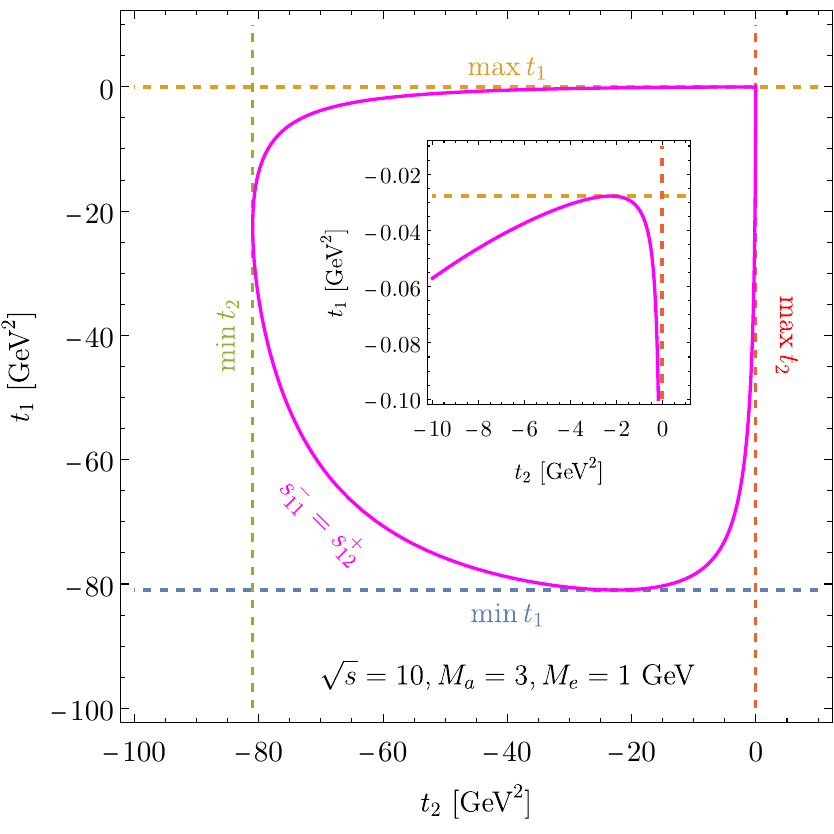}
\caption{The magenta solid curve is the phase space boundary of the $(t_1,t_2)$ cross section distribution. Dashed curves are vertical and horizontal tangents. They delimit the minimum and the maximum allowed to $t_{1,2}$. For the clarity of the picture we chose $s=100 \GeV^2, M_a=3 \GeV, M_e= 1 \GeV$.}
\label{fig:finalt1t2sols}
\end{figure}

Interestingly, from \cref{fig:phi_dt1_dt2} we can see that the phase space privileges small values of $|t_{1,2}|$ if $M_e$ is small even when the matrix element is constant. For similar considerations as \cref{section:dst_dt2_ds1}, the maximum is not attained on the phase space boundary, even though the scalar matrix element pushes it in that direction. We can compute the position of the maximum. Indeed, we can see the argument of the logarithm as a function of the form 

\begin{equation}
\alpha:=\frac{f_1(t_2)-t_2-M_e^2-f_2(t_2)}{f_3(t_1,t_2)-t_2-M_e^2-f_4(t_1,t_2)}.
\end{equation}
Its derivative with respect to $t_1$ is

\begin{equation}
\frac{\dd \alpha}{\dd t_1}=\frac{(M_e^2+t_2-f_1 f_2)(f_3'-f_4')}{(M_e^2+t_2-f_3+f_4)^2},
\end{equation}
where $f'_{3,4}:=\frac{\dd f_{3,4}}{\dd t_1}$. The derivative annihilates if $f_3'=f_4'$. The latter is solved by

\begin{equation}
t_1=x_1:=\frac{M_e(t_2-M_a^2)}{M_a+M_e}.
\end{equation}
Imposing symmetry

\begin{equation}
t_2=x_1 \text{ and } t_1=x_1 \text{ if } t_1=t_2=-M_a M_e,
\end{equation}
The value on the maximum is particularly simple: $\ddv{\Phi_3}{t_1}{t_2}(t_1=t_2=-M_a M_e)=\frac{\pi^2}{4 \sqrt{\lin}} \log \left(\frac{s}{M_a^2}\right)$. We chose to normalize the distributions of \cref{fig:phi_dt1_dt2,fig:dst_dt1_dt2} by the distribution evaluated onto $t_1=t_2=-M_a M_e$to show that between the $\Phi_3$ maximum and the $\sigma_3$ maximum there are orders of magnitude of cross section. \\
All properties listed in this section are plotted in \cref{fig:finalt1t2sols}, where the magenta contour is $\partial{\cal V}_t$. Specifically we can see that ${\cal V}_t$ is convex, that ${\cal V}_t$ is continuous and symmetric under $t_1-t_2$ exchange, and is enclosed among its vertical tangents $t_1=\min, \max t_1$, as in \cref{eq:min_t2_chew_low,eq:max_t2_chew_low}, and its horizontal tangents $t_2=\min, \max t_2$. That \cref{eq:min_t2_chew_low,eq:max_t2_chew_low} are minima and maxima to $t_2$ comes from the fact that they were obtained from ${\cal G}_1$ roots when we were studying $\ddv{\sigma_t}{s_1}{t_2}$, the same ${\cal G}_1$ we have here; that they are minima and maxima to $t_1$ comes from symmetry. We provide here their series expansion in $\frac{M_e}{\sqrt{s}}$:

\begin{equation}
\max t_1 = \max t_2= -\dfrac{M_a^4 M_e^2}{s(s-M_a^2)}\left(1 +{\cal O}\left(\dfrac{M_e^2}{s}\right)\right)\label{eq:max_t12}.
\end{equation}

Lastly, from the zoomed inset of \cref{fig:finalt1t2sols} we can appreciate how the phase space is such that the divergences of the $e^+ e^- \to e^+ e^- a$ scattering is just outside the boundary of the physical phase space. The fact that the $t$ channel can approach the divergence so closely but is protected from it as it does not belong to the phase space, makes the cross section bigger than that of the $s$ channel. Vice versa, applying constraints on final particles can cut away the phase space regions close to the divergences, thus reducing the cross section. Specifically, asking for the final leptons to overcome an energy throshold and not to fly close to the beam pipe has this precise effect: the red upper right corner in \cref{fig:dst_dt1_dt2}, first row, i.e. the $(t_1,t_2)$ region with the biggest $\ddv{\sigma_t}{t_1}{t_2}$, would be cut away. Notice that the same cut would have a marginal effect on the $s$ channel. As is clear from \cref{fig:dist_1d}, left panel, a selection killing the $s$ channel is imposing a minimal final leptons angular distance, which would make $u_1$ big and lose the $\dv{\sigma_s}{u_1}$ maximum.\\

Let us be more specific on the kinematic configuration of the VBF maximizing region. The $(t_1,t_2)$ PS boundary $\partial {\cal V}_t$ is symmetric under $t_1 \leftrightarrow t_2$. The cross section maximizing region lies close to the corner of the quadrant. Here the distance from the origin of the $(t_1,t_2)$ plane is minimized. $\partial {\cal V}_t$ is closest to the $(0,0)$ point onto its intersection with the quadrant bisector. The intersection point has coordinates $(\bar{t_1},\bar{t_2})$ with 

\begin{equation}
\bar{t_1}=\bar{t_2}=-\frac{(M_a M_e)^2}{\sqrt{s}(\sqrt{s}-Ma)}.
\end{equation}
In that point the slope of the tangent to $\partial {\cal V}_t$ is -1, i.e. it is orthogonal to the bisector. There are two proofs of these statements:

\begin{enumerate}
\item In the limiting case $M_e=0$ the $(0,0)$ corner is included in $\partial {\cal V}_t$ and the matrix element diverges hence $\bar{t_1}=\bar{t_2}=0$. Since $\partial {\cal V}_t$ is continuous in $M_e$, by slightly modifying $M_e$ the $(\bar{t_1},\bar{t_2})$ must still be the closest point to the origin.
\item $\partial {\cal V}_t$ is made by two branches, functions of $t_2$. They meet onto $t_2=\min t_2$ and $t_2= \max t_2$, i.e. where the tangent is vertical. Let us start from $\min t_2$ and cycle all around $\partial {\cal V}_t$. On the first branch, i.e. from $\min t_2$ to $\max t_2$, the tangent is horizontal only once. On the second branch, i.e. from $\max t_2$ to $\min t_2$, it is the same. Then one branch is concave and the other one is convex. On one branch the slope of the tangent starts from $+\infty$, has a node and sinks down to $-\infty$, then the other branch starts and the slope meets $-\infty, \ 0, \ +\infty$ in this order. This means that each branch has slope -1 only once. Since you can plug $(\bar{t_1},\bar{t_2})$ in $\partial {\cal V}_t$ and verify the slope is -1 there, $(\bar{t_1},\bar{t_2})$ is the only point with slope -1 on the branch it belongs to.
\end{enumerate}

\subsection{Other cross section distributions from symmetry}
Our scattering enjoys two symmetries: the masses identities $m_1=m_2=m_3=m_5=M_e$ and charge conjugation, coming from the fact that electric charge does not play a role in our matrix element. Therefore if we trade $e^\pm \to e^\mp$ neither the matrix element nor the pahse space should change. Let us call this lepton swap (LS) symmetry and derive some cross section distributions simply from it.

\begin{theorem}[LS symmetry and distribution equivalence]
$~$
\begin{enumerate}
\item $\ddv{\sigma_x}{s_1}{s_2}$ is a symmetric function of $s_1$ and $s_2$ for each $x=s,t$.
\item $\ddv{\sigma_x}{t_1}{t_2}$ is a symmetric function of $t_1$ and $t_2$ for each $x=s,t$.
\item With the same notation $\dv{\sigma_x}{s_1}$ ($\dv{\sigma_x}{t_1}$) has the same functional form of $\dv{\sigma_x}{s_2}$ ($\dv{\sigma_x}{t_2}$) in the sense that if the former is a function depending on $s_1(t_1)$, the latter is the same function depending on $s_2(t_2)$. It is exactly the same function because all masses are sent to themselves.
\item $\ddv{\sigma_x}{s_1}{t_1}$ ($\ddv{\sigma_x}{s_1}{t_2}$) has the same functional form of $\ddv{\sigma_x}{s_2}{t_2}$ ($\ddv{\sigma_x}{s_2}{t_1}$).
\end{enumerate}
\end{theorem}

\noindent
We will only prove the straight points, those in parenthesis can be easily deduced from them. Consider the most generic LS invariant matrix element, ideally depending on all of the gMvs $|{\cal M}_x|^2(s_1,s_2,t_1,t_2)$. The cross section is

\begin{equation}
\sigma_x \propto \int d s_1 d s_2 \int d t_1 d t_2 |{\cal M}_x|^2(s_1,s_2,t_1,t_2) \dfrac{\theta(-\Delta_4)}{\sqrt{-\Delta_4}}=: \int d s_1 d s_2 f_1(s_1, s_2).
\end{equation}

\noindent
Perform a change of variables names

\begin{equation}
\sigma_x \propto \int d s_1 d s_2 \int d t_1 d t_2 |{\cal M}_x|^2(s_2,s_1,t_2,t_1) \dfrac{\theta(-\Delta_4)}{\sqrt{-\Delta_4}}=: \int d s_1 d s_2 f_2(s_1, s_2),
\end{equation}

\noindent
but since the whole integrand is LS invariant

\begin{equation}
f_1(s_1,s_2)=f_1(s_2,s_1)\equiv f_2(s_1,s_2)=f_2(s_2,s_1).
\end{equation}

\noindent
Also

\begin{align}
&F_1(s_2):=\int d s_1 f_1(s_1,s_2)= \int d s_1 f_1(s_2,s_1)=\int d s_2 f_1(s_1,s_2), \label{eq:F1_def}\\
&F_2(s_1):=\int d s_2 f_2(s_1,s_2)= F_1(s_1). \label{eq:F2_def}
\end{align}

\noindent
The first equality in \cref{eq:F1_def} comes from $f_1$ symmetry, the second one is a change of variables names. The equality in \cref{eq:F2_def} comes from the $f_1, f_2$ equivalence. This concludes our proof.\\

It is easy to show that some of the above symmetries can be transposed to the polar coordinate language too as it is sufficient to show that the change of variables from gMvs to polar variables produces a symmetric jacobian. For example the jacobian bringing from $(s_1,s_2)$ to $(E_3,E_5)$ is a constant, or

\begin{align}
&|J((s_2,t_1)\to (E_3,c_3))| = 2 \sqrt{s} \sqrt{s-M_e^2}\sqrt{E_3^2-M_e^2},\\
&|J((s_1,t_2)\to (E_5,c_5))| = 2 \sqrt{s} \sqrt{s-M_e^2}\sqrt{E_5^2-M_e^2},
\end{align}

\noindent
which are transformed one into the other via $E_3 \leftrightarrow E_5$.

Therefore, through LS symmetry we can write

\begin{align}
&
\begin{aligned}
\dv{\sigma_t}{t_1}&= \dfrac{\sigma_0 \pi^2}{4 M_e^2} \dfrac{1}{t_1^2(t_1-M_a^2)^2}\bigg[\dfrac{1}{2}\sqrt{\lambda(s_2^-,M_a^2,M_e^2)}(s_2^--(M_a^2+M_e^2))\\
&+2(M_a M_e)^2 \log\left(\dfrac{s_2^--(M_a^2+M_e^2)-\sqrt{\lambda(s_2^-,M_a^2,M_e^2)}}{2 M_a M_e}\right)\bigg],
\end{aligned}\label{eq:d_sigma_t_dt1}\\
&	\ddv{\sigma_t}{s_2}{t_1}=\dfrac{\sigma_0\pi}{M_e^2(M_a^2-t_1)^2} \sqrt{\lambda(s_2,M_e^2,M_a^2)}\theta(-{\cal G}_1)
\label{eq:dss_ds1_ds2}.
\end{align}

Lastly, by the definition of $u_1$ we can write

\begin{equation}
\ddv{\sigma_s}{s_1}{s_2}=\dfrac{\sigma_0 \pi \sqrt{\lin \Lambda(2 M_e^2+M_a^2+s-s_1-s_2)}}{s^3(2 M_e^2+M_a^2+s-s_1-s_2)^3}
\end{equation}
which is symmetric under $s_1 \leftrightarrow s_2$.

\section{Conclusion}\label{sec:conclusion}
The $2 \to 3$ scattering phase space is foundational to all particle collider experiments, yet its analytic properties have remained largely unexploited in phenomenology. This work closes that gap by providing the first comprehensive analytic treatment that maintains all particle masses non-zero and incorporates non-constant matrix elements - requirements that are unavoidable in realistic Beyond Standard Model searches. Our treatment is universal: it does not depend on topology, kind of collider or kind of interaction, however the necessity of such an analytic grasp is particularly evident for VBF scatterings. Indeed, the typical small scale $M_e/\sqrt{s}$ can lead MC generators to fail. Approximations like EVA are an option, but their applicability is not always guaranteed \cite{eva_fail}. The Gram Determinant formalism provides an exact, closed-form expressions that sometimes can be the only reliable alternative.
In this paper we propose a Lorentz invariant language to describe matrix elements and phase spaces; we recall a dictionary between Gram Determinants and polar coordinates (\cref{signs_and_bounds}); we revisit why null Gram determinants correspond to the boundary of the phase space (\cref{sec:proof_ps_boundary}); we show how the phase space is defined by nested inequalities on Gram Determinants and that lower order constraints define a larger volume phase space that encloses higher order constraints (\cref{corol:phase_space_approx}); we compute some of the most used cross section distributions in particle physics, and provide their physical phase space explicitly keeping all masses non zero and assuming a fully scalar $s-$ and $t-$channel matrix element. We derive them either directly or thanks to the symmetries of the considered scattering. We list them all in  \cref{tab:all_xsec_dist}: the reader dealing with similar processes, even in a theory with spins, can use them as a ready-made tool, with the only need of plugging in the masses of their process.

\begin{table}[!h]
\centering
\begin{tabular}{cccc}
Distribution	& Equation & Approximation & Boundaries\\
\hline
\hline
$\ddv{\sigma_t}{s_1}{t_2}$	&  4.31 & & 4.23, 4.24, 4.25  \\
\hline
$\ddv{\sigma_t}{s_1}{t_2}$	&  4.97 & & 4.23, 4.24, 4.25  \\
\hline
$\frac{\dd \sigma_t}{\dd t_2}$	& 4.39 & 4.40 &4.24, 4.25 \\
\hline
$\frac{\dd \sigma_t}{\dd t_1}$	& 4.96 & 155 &4.24, 4.25 \\
\hline
$\ddv{\sigma_s}{s_1}{u_1}$	& 4.48  &  & 4.49\\
\hline
$\ddv{\sigma_s}{s_1}{s_2}$	& 4.98  &  & 4.49\\
\hline
$\frac{\dd \sigma_s}{\dd u_1}$	& 4.52 &  4.53 & 4.55\\
\hline
$\sigma_s$ & 4.57& & -\\
\hline
$\ddv{\sigma_t}{t_1}{t_2}$ & 4.81 & 4.82 & 4.64-4.70, 4.87
\end{tabular}
\caption{List of all the cross section distributions we computed in this paper.}
\label{tab:all_xsec_dist}
\end{table}

When deriving all these results we left behind many others. Indeed, this is the first of a series of papers: the kinematic phase space is dense with information that need not to be left unexploited. In \cite{gram_4d_lik} we use the property of $\Delta_4$ of being an accumulation point. Indeed the integrable divergence of the phase space density is such that scattering events will preferably accumulate onto the surface $\Delta_4=0$. An overdensity on the curve $\Delta_4=0$ for a specific $M_a$ is indications of New Physics. In \cite{gram_cusps} we propose the \textit{edges} of the phase space as a discovery tool for BSM particles. In \cite{gram_epa} we leverage the simplicity of this Lorentz-invariant language to quantify the errors we make in different approximations like dropping $s-$channel topologies or dropping subdominant regions of the phase space.

\section*{Acknowledgement}
FA warmly thanks Konstantin T. Matchev for their inspiring and encouraging conversation; Andrea Thamm for the great help of all these years; Roberto Franceschini and Diego Redigolo for supervision; Alessio Mastroddi for his friendly support.

\bibliographystyle{JHEP}
\bibliography{bibliography}

\appendix

\section{Definitions}\label{app:definitions}
\subsection{Polar coordinates}
\begin{itemize}
\item $m_i=$ mass of the $i$th particle. For $e^+ e^- \to e^+ e^- a$ we used $m_i=M_e \forall i \neq4, m_4=M_a$.
\item $\epsilon:=\dfrac{M_e}{\sqrt{s}}$
\item $E_i=$ energy of the $i$th particle.
\item $\theta_i=$ polar angle of the $i$th particle.
\item $c_i, s_i=$ cosine and sine of the $i$th particle polar angle respectively.
\item $c_{ij}=$ cosine among the $i$th and $j$th particle.
\item $\phi_i=$ azimuthal angle of the $i$th particle.
\item $c^{\phi}_i,s^{\phi}_i=$ cosine and sine of the $i$th particle azimuthal angle respectively.
\end{itemize}

\subsection{$2 \to 3$ scattering definitions}\label{shortcuts:sec:definitions}

\begin{convention}
\begin{align}
&p_1+p_2 \to p_3+p_4+p_5,\\
&p_i^2=m_i^2,\\
&\vec{m}:=(m_1,m_2,m_3,m_4,m_5).
\end{align}
\end{convention}

\begin{definition}
$\mathbf{{\cal P}_5}=2\to3$ scattering or $1 \to 4$ decay phase space.
\end{definition}

\begin{definition}[gMvs $(s_1,s_2,t_1,t_2)$ and auxiliary variables $(u_{1,5})$] 
\begin{equation}
\begin{split}
& s:= (p_1+p_2)^2,\\
& s_1:=(p_3+p_4)^2,\\
& s_2:= (p_4+p_5)^2,\\
& t_1:= (p_1-p_3)^2,\\
& t_2:=(p_2-p_5)^2,
\end{split}
\qquad \qquad
\begin{split}
& u_1:=(p_3+p_5)^2,\\
& u_2:=(p_1-p_5)^2,\\
& u_3:=(p_2-p_3)^2,\\
& u_4:=(p_2-p_4)^2,\\
& u_5:=(p_1-p_4)^2.
\end{split}
\end{equation}
\end{definition}

\begin{property}[From scalar products to gMvs]
\begin{equation}
\begin{split}
& p_{12}= \dfrac{s- m_1^2-m_2^2}{2},\\
& p_{34}= \dfrac{s_1- m_3^2-m_4^2}{2},\\
& p_{45}= \dfrac{s_2- m_4^2-m_5^2}{2},\\
& p_{13}= -\dfrac{t_1- m_1^2-m_3^2}{2},\\
& p_{25}= -\dfrac{t_2- m_2^2-m_5^2}{2},
\end{split}
\qquad \qquad
\begin{split}
& p_{23}= \dfrac{s-s_2+t_1-m_1^2}{2},\\
& p_{24}= \dfrac{s_2 -t_1+t_2-m_5^2}{2},\\
& p_{14}= \dfrac{s_1 + t_1-t_2-m_3^2}{2},\\ 
& p_{15}= \dfrac{s-s_1 +t_2-m_2^2}{2},\\
& p_{35}= \dfrac{s -t_1+t_2-m_5^2}{2}.
\end{split}
\end{equation}
\end{property}

\begin{property}[From gMvs to auxiliary and vice versa] 

\begin{equation}
\begin{split}
& u_1=m_3^2+m_4^2+m_5^2+s-s_1-s_2,\\
& u_2=m_1^2+m_2^2+m_5^2-s+s_1-t_2,\\
& u_3=m_1^2+m_2^2+m_3^2-s+s_2-t_1,\\
& u_4=m_2^2+m_4^2+m_5^2-s_2+t_1-t_2,\\
& u_5=m_1^2+m_3^2+m_4^2-s_1-t_1+t_2,
\end{split}
\qquad \qquad
\begin{split}
&s=u_1-u_4-u_5+m_1^2+m_2^2+m_4^2,\\
&s_1=u_2-u_3-u_4+m_2^2+m_3^2+m_4^2,\\
&s_2=-u_2+u_3-u_5+m_1^2+m_4^2+m_5^2,\\
&t_1=-u_1-u_2+u_4+m_1^2+m_3^2+m_5^2,\\
&t_2=-u_1-u_3+u_5+m_2^2+m_3^2+m_5^2.
\end{split}\label{eqs:from_mad_to_aux_and_back}
\end{equation}
\end{property}

\begin{property}[Scalar Matrix elements]
\begin{equation}
{\cal M}_s \to \dfrac{1}{s u_1}, \qquad {\cal M}_t \to \dfrac{1}{t_1 t_2}.
\end{equation}
\end{property}

\begin{definition}[${\cal G}_{1,2}$]
\begin{align}
& {\cal G}_1=\dfrac{M_e^2 s_1^2}{2}-s_1 \left(\dfrac{s t_2}{2}+M_e^4\right)+ \dfrac{1}{2}(M_e^6-3 M_e^2 s t_2 + s t_2(s+t_2))\\
& \begin{aligned}
{\cal G}_2&=\dfrac{t_1 s_1^2}{2}-\dfrac{1}{2}s_1 t_1(M_a^2+2 M_e^2+t_2-t_1)\\
&+ \dfrac{1}{2}(M_a^4 M_e^2+M_e^4 t_1+ M_e^2 t_2(t_2-t_1) + M_a^2(t_1 t_2-M_e^2(t_1+2 t_2)))
\end{aligned}
\end{align}
\end{definition}

\begin{definition}
In a scattering where the two incoming four-momenta are $p_{1,2}$, the collision flux will always depend on a triangular lambda like
\begin{equation}
\lin:=\lambda(s,m_1^2,m_2^2).
\end{equation}
\end{definition}

\section{Byers and Yang}\label{sec:byers_yang}

Results of the previous sections are special cases of Byers and Yang mathematical description of scattering and decays phase spaces \cite{byers_yang}. In this section I will walk in their footsteps in a more pictorial, physicists friendly way, as it will be helpful to better understand the $2 \to 3$ scattering.\\
They base their beautiful review on the space ${\cal S}_n$ of all the scalar products (\cref{def:m_n}) of the considered process (assumption \ref{assu:masses}) and associate a Gram Matrix to each point of said space. Many remarks can be made on these real matrices, especially on their eigenvalues. The sign of the eigenvalues of these matrices can be used to classify subspaces of the entire space. The key question one can pose on ${\cal S}_n$ is: is there a transformation that can make me continuously go from one point of ${\cal S}_n$ to another? The answer is: it depends (\cref{theo:r_connect}). If we want to stick to the physical case (1+3 dimensional space, \cref{theo:d_1_3}), we need to remember that to transform from forward to backward time-like fourmomenta there is no continuous transformation. This causes the 1+3 phase space to break down in pieces (\cref{theo:r_disconnect}), that Byers and Yang analyse. They finally provide the GD phase space boundary theorem and their Kibble curve derivation.

\bigskip

\begin{assumption}[Positive masses]\label{assu:masses}
We will consider $1 \to n$ decays or $2 \to n$ scatterings with non-zero masses:

\begin{equation}
m_i\neq0 \ \forall i=1, \dots n.
\end{equation}
\end{assumption}

\begin{definition}[Generalized Minkowskian space $D_{ab}$ and Generalized Minkowskian Metric $\eta^{(ab)}$]
In a generalized $d$ dimensional space $D_{a b}$ we pick $a$ dimensions to be time dimensions and $b$ to be space, $a+b=d$. In particular $D_{13}$ is the physical Minkowskian space $\mathbb{M}_4$. Correspondingly, the usual mostly negative Minkowskian metric is expressed in terms of the block matrices
\begin{equation}
\eta^{(a b)}:=
\begin{pmatrix}
\mathbb{1}_a & 0\\
0 & -\mathbb{1}_b
\end{pmatrix},
\end{equation}

\noindent
where $\mathbb{1}_d$ indicates a unit $d \times d$ matrix.
\end{definition}

\begin{definition}[Scalar product space ${\cal S}_n$]\label{def:m_n}
${\cal S}_n$ is the space of all scalar products one can make with $n$ fourmomenta. To each point in ${\cal S}_n$ corresponds a Gram Matrix and a Gram Determinant.
\end{definition}

\begin{definition}[Scalar product space region $r_{a_- a_+}$]
Let $M_n$ be a generic Gram Matrix of dimension $n$. Its ordered eigenvalues are $\{\lambda_i\}_{i=1}^n \in \mathbb{R}$. We say $M_n$ belongs to the region $r_{a_- a_+}$ if the number of $M_n$ positive eigenvalues is $a_+$ and the number of negative eigenvalues is $a_-$. We can define the number of null eigenvalues as $a_0=n-a_--a_+$.
\end{definition}

\begin{corollary}[Regions and independent fourmomenta]
If $M_n \in r_{a_- a_+}$ then exactly $a_-+a_+$ fourmomenta are independent, in fact the other ones give null eigenvalues.
\end{corollary}

\begin{theorem}[Space and time dimension of an $M_n$ realization: from $r_{a_- a_+}$ to fourmomenta]
If $M_n \in r_{a_- a_+}$ there exists a physical configuration of $n$ fourmomenta in dimension $D_{a_- a_+}$ making up $M_n$, of which only $a_-+ a_+$ are independent. Differently stated, there is a realization for $M_n$. $a_-$ is the smallest possible space dimension, $a_+$ is the smallest time dimension, meaning that it is always possible to embed the realization in a bigger space but there is no transformation that can be operated to make some space or time dimension redundant.
\end{theorem}

\begin{theorem}[From fourmomenta to $r_{a_- a_+}$]
The converse is true: if $n$ fourmomenta live in $D_{a- a_+}$ and $a_-+a_+$ of them are independent, then the matrix $M_n$ made from those fourmomenta lives in $r_{a_- a_+}$.
\end{theorem}

\begin{theorem}[$D_{13}$]\label{theo:d_1_3}
In $D_{13}$ (Minkowski) we can only have  $a_+=0,1$. Since 

\begin{equation}
\tr M_n =\sum_i m_i^2>0,
\end{equation}

\noindent
there must at least be one positive eigenvalue, hence $a_+=1$. This is because the trace of a matrix is the sum of its eigenvalues, therefore, if all the eigenvalues were negative, there would be no way to make the trace positive.\\
The possible regions of ${\cal s}_n$ in $D_{13}$ are

\begin{equation}
r_{13}, \ r_{12}, \ r_{11}, \ r_{10}.
\end{equation}

\end{theorem}

\begin{theorem}[If $a_+>1, \ r_{a_- a_+}$ is connected]\label{theo:r_connect}
Starting from a set of vectors I can always perform a continuous deformation leaving the number of independent fourmomenta and the length of the fourmomenta unchanged.
\end{theorem}

\begin{theorem}[If $a_+=1, \ r_{a_- a_+}$ is made of $2^{n-1}$ connected regions]\label{theo:r_disconnect}
This is because there is no way of continuously transform from forward to backward time-like fourmomenta and vice-versa. However an overall sign will not change $M_n$ hence the regions are $2^{n-1}$.
\end{theorem}

Let us now stick to the $a_+=1$ case and consider a $1 \to n$ decay, so that all final fourmomenta are forward timelike.
There exists a region $r_{1 b}^0$ with all positive scalar products, exactly the one that we carved out in \cref{subsec:branch_selection}.

\begin{definition}[Byers and Yang GDs]
Given a set of $n$ fourmomenta $\{p_i\}_{i=1}^n$, the GDs à la Byers and Yang are

\begin{equation}
\delta_m = \sum_{\vec{p}} \Delta_m(\vec{p}),
\end{equation}

\noindent
where $\Delta_m$ follows \cref{def:gd} and we summed over all possible subset $\vec{p}$ of size $m$ of the set $\{p_i\}_{i=1}^n$.

\end{definition}

\begin{theorem}[Phase Space boundary]
A necessary and sufficient condition for a point of ${\cal S}_n$ to be in $r_{1s}$ is 

\begin{empheq}[left={(-1)^{n+1} \delta_n\empheqlbrace}]{align}
\geq 0 &\text{ if } n<=s+1 \label{eq:positive_by_gds},\\
=0 &\text{ if } n>s+1 \label{eq:null_by_gds}.
\end{empheq}

\noindent
The sufficiency can be proved as follows. Let us write down the characteristic polynomial of $M_n$ as a function of $\lambda$:

\begin{equation}
\lambda^{n-s-1}(\lambda^{s+1}- \delta_1 \lambda^s- \dots \delta_{s+1})=0,
\end{equation}

\noindent
where we stopped at $\delta_{s+1}$ thanks to \cref{eq:null_by_gds}. But then $M_n$ has $n-s-1$ null eigenvalues. Moreover the term in brackets has only 1 change of sign henceforth there only is 1 positive eigenvalue.\\		

For the necessity we need to assume that $M_s$ is in $r_{1s}$ and recover \cref{eq:positive_by_gds,eq:null_by_gds}. \Cref{eq:null_by_gds} is easy because null eigenvalues will make all minors of dimension bigger than $s+1$ null. As for \cref{eq:positive_by_gds}, it is proven via linear independence and masses positivity.
\end{theorem}

\bigskip

Specifically in 3 spatial dimensions \cref{eq:positive_by_gds,eq:null_by_gds} reduce to the discussed Kibble problem.\\ 
$\Delta_3=0$ is an algebraic surface $S$ of degree 3, i.e. it is a cubic surface in $s,t,u$. It has 4 conical points. A conical point of a surface is a point M such that there exists a cone with vertex M tangent at M to the surface. They are

\begin{equation}
\begin{aligned}
& C_0:(x, y, z)=\left(m_2 m_3, m_3 m_1, m_1 m_2\right), \\
& C_1:(x, y, z)=\left(m_2 m_3,-m_3 m_1,-m_1 m_2\right), \\
& C_2:(x, y, z)=\left(-m_2 m_3, m_3 m_1,-m_1 m_2\right), \\
& C_3:(x, y, z)=\left(-m_2 m_3,-m_3 m_1, m_1 m_2\right).
\end{aligned}
\end{equation}

\noindent
$S$ is made by a central piece $S_0$ and 4 horns $H_{0,1,2,3}$ stemming out of the respective $C_i$. $S_0$ is a closed surface contained in the parallelepiped whose vertices are

\begin{equation}
(\pm m_2 m_2, \pm m_3 m_1, \pm m_1 m_2).
\end{equation}

What are the possible regions? In principle all $r_{ab}$ combinations with $a+b=3$ are possible, however some are not allowed by theorem \cref{theo:d_1_3}. They are

\begin{itemize}
\item $r_{30}$, the inside of $S_0$.
\item $r_{20}$, the surface of $S_0$, $C_i$ excluded.
\item $r_{21}$, the region outside $S_0$ and outside the horns.
\end{itemize}

\noindent
Instead the allowed ones are

\begin{itemize}
\item $r_{12}$, the inside of the horns.
\item $r_{11}$, the surfaces of the horns, $C_i$ excluded.
\item $r_{10}$, the $C_i$.
\end{itemize}

\noindent
The Kibble curve is obtained by intersecting the $H_0$ horn with the plane coming from

\begin{equation}
s+t+u=k.
\end{equation}

These properties can be generalized to $1 \to n$ decays and $2 \to n$ scatterings.

\section{$\ddv{\sigma_t}{t_1}{t_2}$ distribution plots}\label{appendix:gallery}

\begin{figure}
\centering
\includegraphics[width=\linewidth]{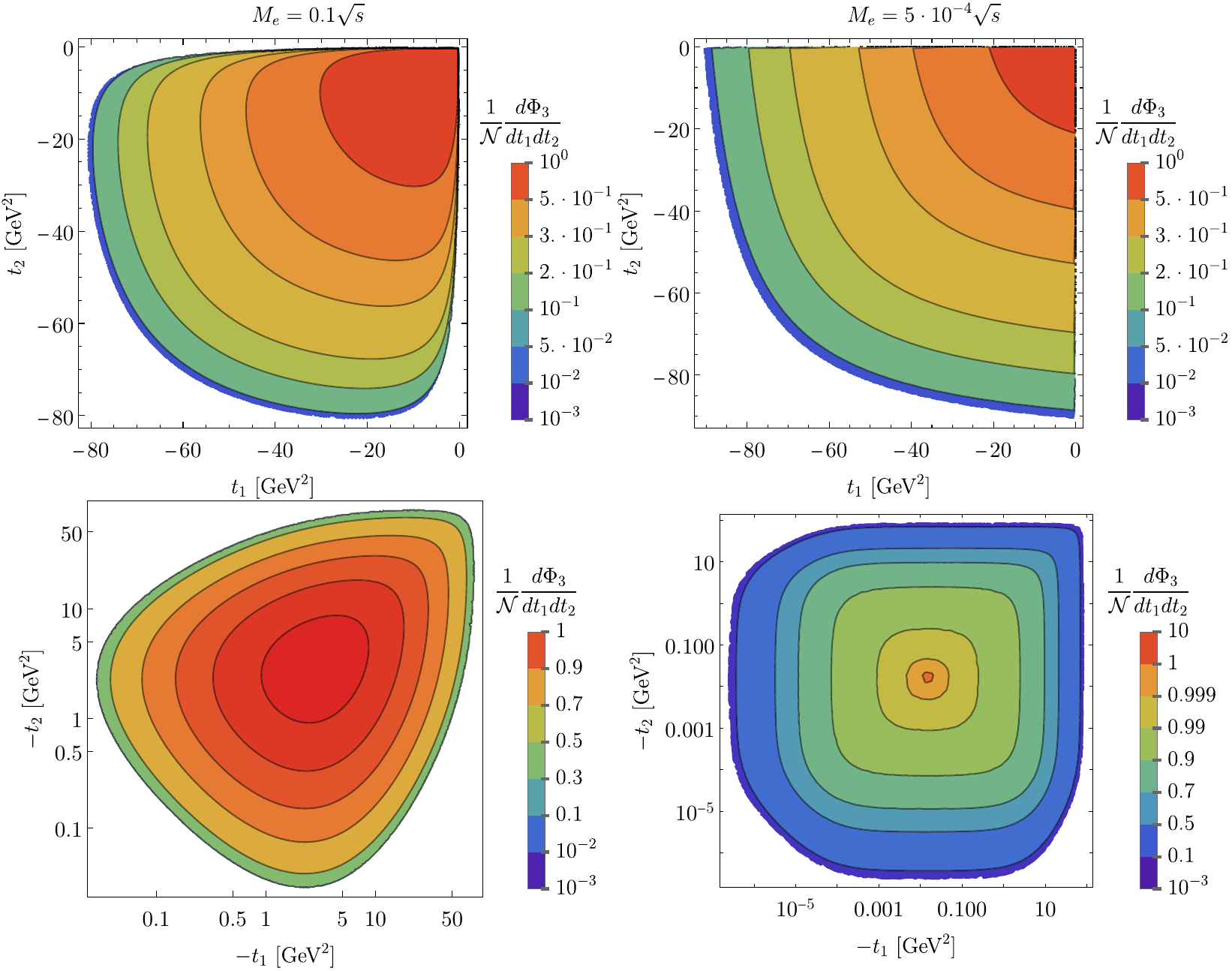}
\caption{Phase space distribution for \cref{eq:ds_dt1_dt2} with constant matrix element. All the plots have $\sqrt{s}=10, M_a=3$ GeV and are rescaled by the norm ${\cal N}:= \ddv{\Phi}{t_1}{t_2} \left(t_1=t_2=-M_a M_e\right)$. The left column has $M_e=0.1 \sqrt{s}$. The right column has $M_e=5 \cdot 10^{-4} \sqrt{s}$.}
\label{fig:phi_dt1_dt2}
\end{figure}

\begin{figure}
	\centering
	\includegraphics[width=\linewidth]{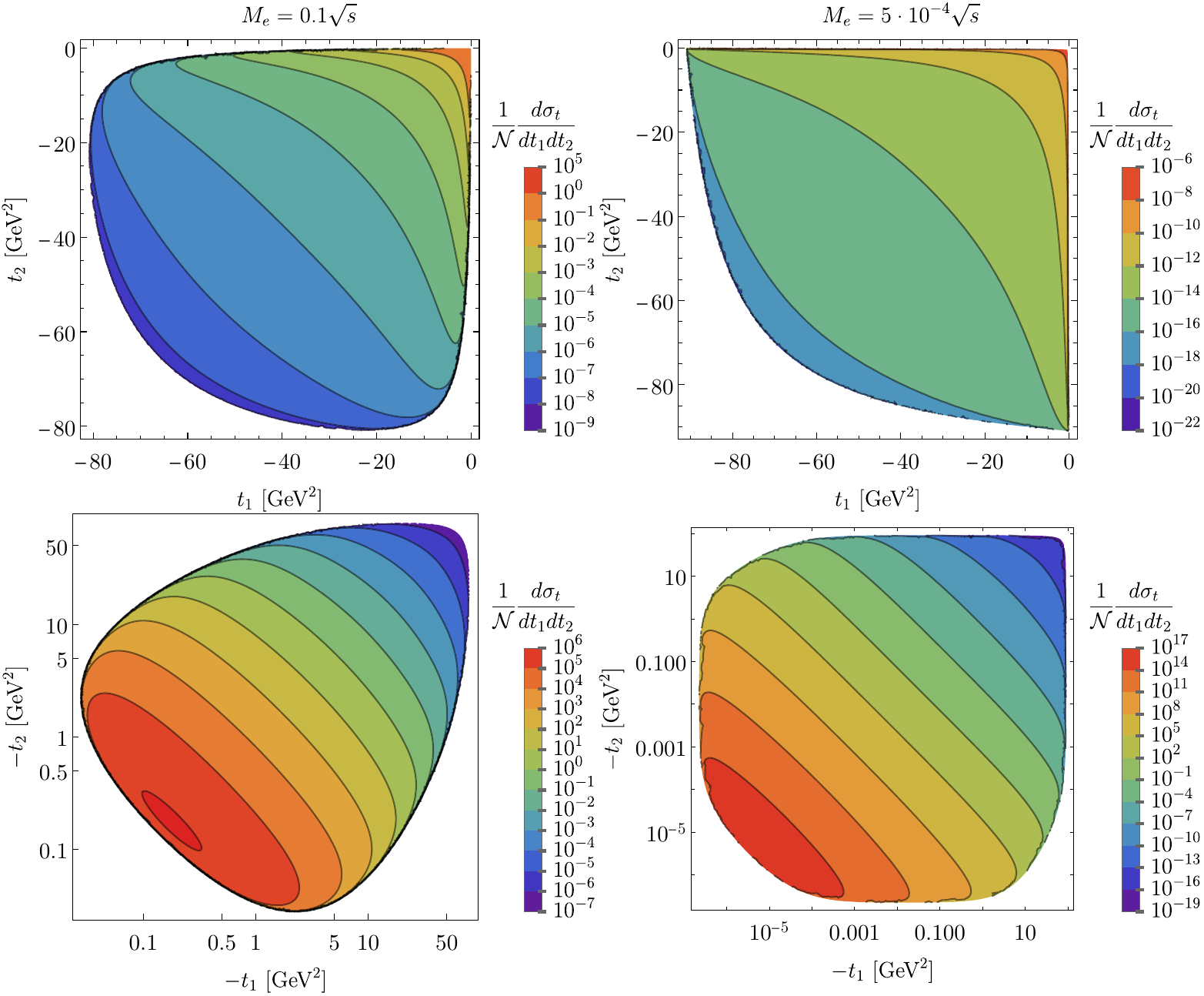}
	\caption{Phase space distribution for \cref{eq:ds_dt1_dt2} with $t-$channelmatrix element. All the plots have $\sqrt{s}=10, M_a=3$ GeV and are rescaled by the norm ${\cal N}:= \ddv{\sigma_t}{t_1}{t_2} \left(t_1=t_2=-M_a M_e\right)$. The left column has $M_e=0.1 \sqrt{s}$. The right column has $M_e=5 \cdot 10^{-4} \sqrt{s}$.}
	\label{fig:dst_dt1_dt2}
\end{figure}

\end{document}